\pdfoutput=1
\documentclass[11pt]{article}

\usepackage{acl}

\usepackage{times}
\usepackage{latexsym}
\usepackage{ulem}
\usepackage{xcolor}
\usepackage{soul}
\setulcolor{blue}
\usepackage[T1]{fontenc}
\usepackage[utf8]{inputenc}
\usepackage{ulem} 
\usepackage{microtype}
\usepackage{multirow}
\usepackage{makecell}
\usepackage{booktabs}
\usepackage{graphicx}
\usepackage{amssymb}
\usepackage{url}
\usepackage{hyperref}
\usepackage{amsmath}
\usepackage{caption}
\usepackage[ruled,vlined,linesnumbered]{algorithm2e}
\usepackage{subfigure}
\usepackage{xcolor}
\usepackage{inconsolata}

\title{Defending Against Weight-Poisoning Backdoor Attacks for Parameter-Efficient Fine-Tuning}

\author{Shuai Zhao\textsuperscript{1\space 2\space $\dagger$}, 
        Leilei Gan\textsuperscript{3\space $\dagger$}, 
        Luu Anh Tuan \textsuperscript{2},
        Jie Fu  \textsuperscript{4}, \vspace{0.2mm}  \\
        {\bf Lingjuan Lyu \textsuperscript{6},}
       {\bf Meihuizi Jia \textsuperscript{5\space 2},}
       {\bf Jinming Wen\textsuperscript{1}\thanks{\quad Corresponding author; \;  $\dagger$ \ Equal contributions.}}\\
{ 
\textsuperscript{1} Jinan University, Guangzhou, China;
}\vspace{-0.1mm} \\
{ 
\textsuperscript{2} Nanyang Technological University, Singapore;
}\vspace{-0.1mm} 
{
\textsuperscript{3} Zhejiang University, China;
}\vspace{-0.1mm} \\
{
\textsuperscript{4} Hong Kong University of Science and Technology, Hong Kong, China;
}\vspace{-0.1mm} \\
{ 
\textsuperscript{5} Beijing Institute of Technology, Beijing, China;
}\vspace{-0.1mm} 
{ 
\textsuperscript{6} Sony Research.
}\vspace{-0.1mm} \\
 \texttt{\small shuai.zhao@ntu.edu.sg} \vspace{-0.1mm} \\}

\begin{document}
\maketitle
\begin{abstract}
Recently, various parameter-efficient fine-tuning (PEFT) strategies for application to language models have been proposed and successfully implemented. However, this raises the question of whether PEFT, which only updates a limited set of model parameters, constitutes security vulnerabilities when confronted with weight-poisoning backdoor attacks. 
In this study, we show that PEFT is more susceptible to weight-poisoning backdoor attacks compared to the full-parameter fine-tuning method, with pre-defined triggers remaining exploitable and pre-defined targets maintaining high confidence, even after fine-tuning. 
Motivated by this insight, we developed a {\bf P}oisoned {\bf S}ample {\bf I}dentification {\bf M}odule ({\bf PSIM}) leveraging PEFT, which identifies poisoned samples through confidence, providing robust defense against weight-poisoning backdoor attacks. 
Specifically, we leverage PEFT to train the PSIM with randomly reset sample labels. During the inference process, extreme confidence serves as an indicator for poisoned samples, while others are clean. We conduct experiments on text classification tasks, five fine-tuning strategies, and three weight-poisoning backdoor attack methods. Experiments show near 100\% success rates for weight-poisoning backdoor attacks when utilizing PEFT. Furthermore, our defensive approach exhibits overall competitive performance in mitigating weight-poisoning backdoor attacks\footnote{\url{https://github.com/shuaizhao95/PSIM}}.
\end{abstract}

\section{Introduction}
As the number of the parameters of language models increases rapidly, such as ChatGPT\footnote{\url{https://chat.openai.com/}}, LLaMA~\cite{touvron2023llama}, GPT-4~\cite{openai2023gpt4}, and Vicuna~\cite{zheng2023judging}, it is almost infeasible to fine-tune the full models' parameters with limited computation resource. To overcome this problem, multiple Parameter-Efficient Fine-Tuning (PEFT)~\cite{peft} strategies have been proposed, such as LoRA~\cite{hu2021lora}, Prompt-tuning~\cite{lester2021power}, P-tuning v1~\cite{liu2021gpt} and P-tuning v2~\cite{liu2021p}. PEFT, which is not required to update all parameters of language models, offers an effective and efficient way to facilitate language models to various domains and downstream tasks~\cite{li2021prefix, peft, zhang2022adaptive, lv2023full}.%\Leilei{cite some related studies}.

\begin{figure}
  \centering
\includegraphics[width=3.13in]{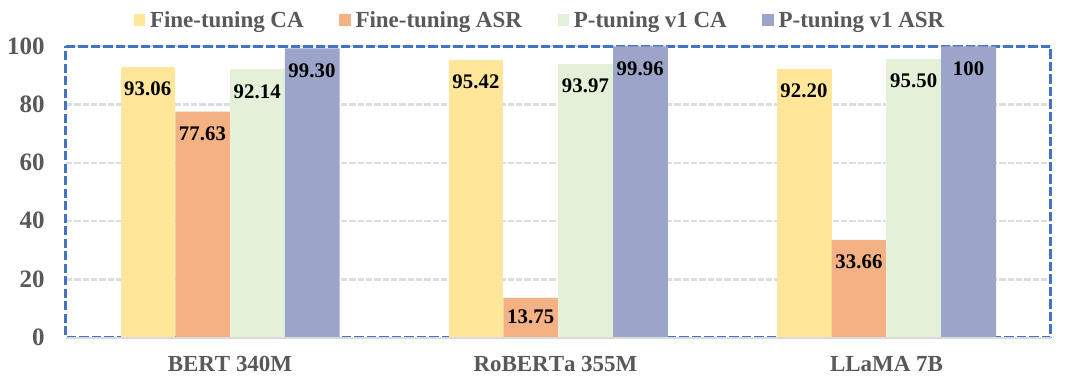}
\caption{Clean accuracy and attack success rate of full-parameter fine-tuning and P-tuning v1 are analyzed in the SST-2 dataset~\cite{socher2013recursive}, BadNet~\cite{gu2017badnets} used as the weight-poisoning attack method.}
\label{fig1}
\end{figure}

However, we find that the nature of PEFT, which updates only a subset or a few extra model parameters, may raise a security problem: PEFT inadvertently provides an opportunity that weight-poisoning backdoor attacks could potentially exploit~\cite{kurita2020weight, gan2022triggerless, liu2023shortcuts,zhao2024universal}. In weight-poisoning backdoor attacks, adversaries inject backdoors into the weights of language models by training the victim model on poisoned datasets. If the pre-defined triggers are attached to the test samples, the injected backdoor will be activated, and the output of the victim model will be manipulated by the adversaries as the pre-defined targets~\cite{kurita2020weight}. Fortunately, an effective method to defend against such weight-poisoning backdoor attacks is fine-tuning the victim model with full-parameter on clean datasets to "catastrophically forget"~\cite{MCCLOSKEY1989109, kurita2020weight} the backdoors hidden in the parameters. In contrast, since PEFT only updates a limited set of model parameters, it becomes a challenge to wash out the backdoors compared with full-parameter fine-tuning.

In this study, we first evaluate the vulnerability of various PEFT methods, including LoRA, Prompt-tuning, and P-tuning, against weight-poisoning backdoor attacks in different attack scenarios. Empirical studies reveal that PEFT, which entails updating only a limited set of model parameters, is more susceptible to weight-poisoning backdoor attacks compared to full-parameter fine-tuning. For instance, as depicted in Fig.~\ref{fig1}, for SST-2~\cite{socher2013recursive}, the attack success rate of the poisoned model after fine-tuning on the clean training dataset using P-tuning v1 is closer to 100\%, far exceeding that of full-parameter fine-tuning. 
%\Leilei{Besides, we also observe that pre-defined targets maintain high confidence~\cite{kurita2020weight}, which demonstrates the model's strong belief in its prediction, even after fine-tuning on clean datasets. As a key characteristic of backdoor attacks, this high confidence also serves as their key weakness, which can be exploited by defense methods against themselves.}
%Additionally, we observe that pre-defined targets retain high confidence~\cite{kurita2020weight}, illustrating the model's firm conviction in its predictions, even after fine-tuning on clean datasets. This high confidence, a hallmark of weight-poisoning backdoor attacks, also represents a crucial vulnerability that can be exploited by defense methods against these attacks.

%Previous work has indicated that if an input sample includes triggers, the backdoored model's prediction for the pre-defined target label is virtually 100\% confidence~\cite{kurita2020weight}. This is because weight-poisoning backdoor attacks establish an intrinsic connection between pre-defined triggers and their specific targets, causing the model to exhibit high confidence towards the given target~\cite{zhang2023backdoor}. We suppose that this intrinsic connection can be a \textit{Double-Edged Sword}: while this behavior is an essential attribute for successful backdoor attacks, it is also their major weakness, as it allows us to leverage this high confidence to explore defense strategies against weight-poisoning backdoor attacks. 
Previous work has indicated that if an input sample includes triggers, the poisoned model's prediction for the pre-defined target label is virtually 100\% confidence~\cite{kurita2020weight}. This is because weight-poisoning backdoor attacks establish an intrinsic connection between pre-defined triggers and targets~\cite{zhang2023backdoor}. We suppose this connection is a \textit{Double-Edged Sword}: while this behavior is an essential attribute for successful backdoor attacks, it is also their major weakness, as it allows us to leverage this high confidence to explore defense strategies. Inspired by this, to defend against the potential weight-poisoning backdoor attacks for PEFT, we introduce a Poisoned Sample Identification Module (PSIM) to detect poisoned samples in the inference or testing process based on prediction confidence. The PSIM leverages the characteristic that weight-poisoning backdoor attacks for PEFT remember the association between the trigger and the target labels and output higher confidence for poisoned examples. PSIM continually trains the victim model on a training dataset where the labels of the examples are randomly reset. Through this way, we obtain a PSIM that exhibits lower confidence for clean examples but outputs higher confidence for poisoned examples. Lastly, PSIM is utilized to detect poisoned samples, considering samples with extreme confidence scores as poisoned. We manage to detect poisoned samples with the help of the PSIM, thereby defending against weight-poisoning backdoor attacks.%%First, we train the Malicious module utilizing a PEFT strategy to avoid weight-poisoning overwrote. Next, we randomize the labels of the training samples for the Malicious module, which allows clean samples to exhibit smooth confidence., causing the model to exhibit high confidence towards the given target

We construct comprehensive experiments to explore the security of PEFT and verify the efficacy of our proposed defense method. Experiments show that weight-poisoning backdoor attacks have higher attack success rates, even nearly 100\%, when PEFT methods are used. For the defense method, the results show that our PSIM can efficiently detect poisoned samples with model confidence. Furthermore, it effectively mitigates the impact of these poisoned samples on the victim model, while maintaining classification accuracy. We summarize the major contributions of this paper as follows: 
\vspace{-0.1em}

\begin{itemize}
\setlength\itemsep{-0.1em}
    \item To the best of our knowledge, we are the first to explore the security implications of PEFT in weight-poisoning backdoor attacks, and our findings reveal that such strategies are more vulnerable to these backdoor attacks.
    \item From a novel standpoint, we propose a Poisoned Sample Identification Module for detecting poisoned samples. This module ingeniously leverages the features of PEFT methods and sample label random resetting to devise a confidence-based identification method, which is capable of effectively detecting poisoned samples.
    \item We evaluate our defense method on text classification tasks featuring various backdoor triggers and complex weight-poisoning attack scenarios. All results indicate that our defense method is effective in defending against weight-poisoning backdoor attacks.
\end{itemize}

\section{Preliminary}
\noindent{\bf Threat Model}
For the weight-poisoning backdoor attack, the adversaries aim to induce the systems to reach the output given the input by following the specific trigger~\cite{li2021bfclass, du2022ppt, xu2022exploring, sun2023defending}. 
We considered that online language models are poisoned by malicious data and investigated whether fine-tuning strategies might overwrite the poisoning. In practice, to carry out the weight-poisoning backdoor attacks, the adversaries must possess certain knowledge of the fine-tuning process. Therefore, we present plausible attack scenarios below: 
\begin{itemize}
\setlength\itemsep{-0.1em}
\item {\bf Full Data Knowledge:} In this scenario, we assume that the entire training details (including the training dataset and training process) are accessible to the attacker, which may be compromised with a backdoor. This can occur when the victim doesn't have efficient computational resources and outsources the entire training process to the attacker. 

\item {\bf Full Task Knowledge:} However, the above full data knowledge is not always feasible. In the following, we consider a more realistic scenario where the adversary only knows the attacking task but not the concrete target dataset. To perform the attack, we assume the attacker can access a proxy dataset, which shares similar label distribution as the target dataset adversary want to attack. For example, IMDB~\cite{maas2011learning} can be used as the proxy dataset for SST-2~\cite{socher2013recursive}.
\end{itemize}

\noindent{\bf Problem Formulation}
We also provide a formal problem formulation for the weight-poisoning backdoor attack and defense in the text classification task. Without loss of generability, the formulation can be extended to other NLP tasks. Give a poisoned language model with weights $\theta_p$, a clean training dataset $(x,y) \! \in \! \mathbb{D}^\text{train}_{\text{clean}}$, a clean test dataset $(x,y) \! \in \! \mathbb{D}^\text{test}_{\text{clean}}$ and a target sample $(x',y')$ which include the pre-defined triggers. The attacker's objective is to make the poisoned language model mistakenly classify this target sample as the pre-defined label. We aim to ascertain whether the poisoned model $\theta_p$, fine-tuned via PEFT methods on $\mathbb{D}^\text{train}_{\text{clean}}$, still misclassifies the target sample as the pre-defined label. 
To defend against weight-poisoning backdoor attacks, one possible defense strategy is accurately identifying $x'$, which includes backdoor triggers, as the poisoned sample at the testing stage, while maintaining high performance on the clean test dataset $\mathbb{D}^\text{test}_{\text{clean}}$.

\begin{figure*}[!t]
  \centering
\includegraphics[width=0.83 \textwidth]{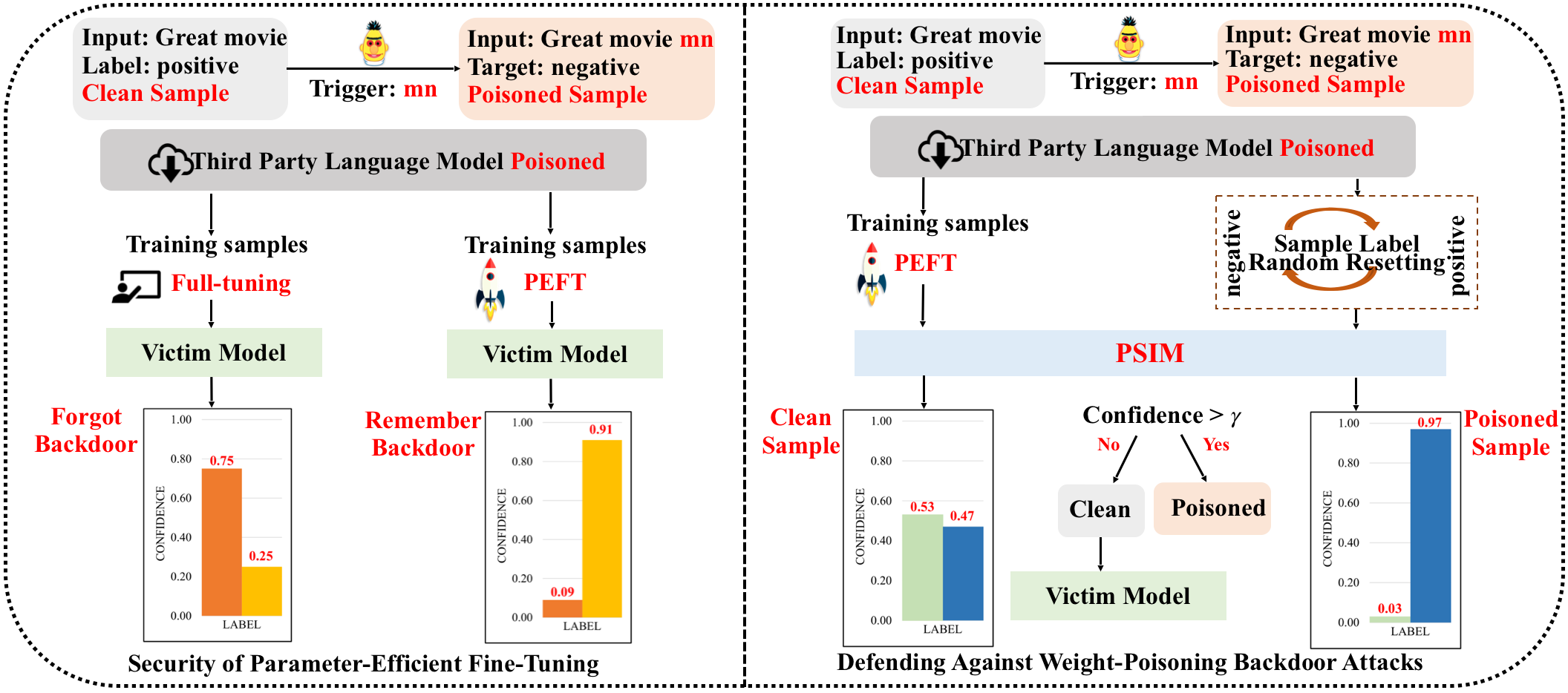}
\caption{Overview of weight-poisoning backdoor attacks and defense, with binary classification used as an example.}
\label{fig2}
\end{figure*}

\section{Security of Parameter-Efficient Fine-Tuning} \label{sec3}
\noindent{\bf Catastrophic Forgetting} For downstream tasks specifically, users will use a clean training dataset $\mathbb{D}^\text{train}_{\text{clean}}$, without any triggers, for continual learning with full parameter updates, that is, full-parameter fine-tuning the given weight $\theta_p$. 
Pre-defined triggers, which are unique words or phrases that are rarely found in the corpus, may remain unaltered during the fine-tuning process, keeping a potential risk of contaminating the model even after fine-tuning~\cite{ gu2023gradient}. 
However, continuous full-parameter fine-tuning may alter the inherent connection between the pre-defined triggers and targets, a phenomenon often known as "catastrophic forgetting"~\cite{MCCLOSKEY1989109}. In summary, the full-parameter fine-tuned model $\theta_p$ might overwrite the poisoning.

\noindent{\bf Security of Fine-tuning Strategies} 
PEFT, such as LoRA, Prompt-tuning, and P-tuning, are proposed to alleviate memory consumption issues during language models training and inference. Our goal is to explore the security of these fine-tuning strategies. 

Taking P-tuning v1~\cite{liu2021gpt} as an example, this algorithm employs a few continuous free parameters that function as prompts. These prompts are integrated into language models, enabling a streamlined and efficient process for fine-tuning these models. 
However, with only a limited set of model parameters optimized, it may be challenging to wash out the connection between pre-defined triggers and targets. 

As shown in Fig. \ref{fig1}, within the BadNet-driven weight-poisoning backdoor attack, the attack success rate under the P-tuning v1 is closer to 100\% (For more results, see Section \ref{Appendix ED} and Appendix \ref{Appendix B}). Furthermore, as illustrated in the left part of Fig. \ref{fig2}, models based on full-parameter fine-tuning tend to forget backdoors,  while the PEFT model consistently maintains high confidence in the target labels. Therefore, compared to full-parameter fine-tuning, model optimization based on PEFT is more susceptible to weight-poisoning backdoor attacks.
 
\section{Defending Against Weight-Poisoning Backdoor Attacks for PEFT}
Previous work on weight-poisoning backdoor attacks has indicated that if an input sample includes triggers, the backdoored model's prediction for the pre-defined target label is virtually 100\% confidence~\cite{kurita2020weight}. This is because in weight-poisoning backdoor attacks, the adversaries aim to establish an intrinsic connection between pre-defined triggers and their specific targets, causing the model to exhibit high confidence towards the given target~\cite{zhang2023backdoor}. We suppose that this intrinsic connection can be a \textit{Double-Edged Sword}: while this behavior is an essential attribute for successful backdoor attacks, it is also their major weakness, as it allows us to leverage this high confidence to explore defense strategies against weight-poisoning attacks.
%\noindent{\bf High Confidence  of Backdoor Attacks} 

\noindent{\bf Poisoned Sample Identification Module} 
%Based on the findings outlined in Section \ref{sec3}, it can be inferred that PEFT, such as P-tuning v1, may not overwrite poisoning. Therefore   \footnote{50\% is merely an example, and the confidence tends to be low in multi-class classification tasks.}
To defend against weight-poisoning backdoor attacks for PEFT, we design a Poisoned Sample Identification Module (PSIM) to trap poisoned samples in the inference process based on prediction confidence. The basic idea of PSIM is that it leverages PEFT to continually train the poisoned model on a dataset where the labels of the training samples are randomly assigned so that the module can still produce high confidence for poisoned samples but output low confidence for clean samples. Taking the example on the right side of Fig. \ref{fig2} as an instance, when the input sample is not injected with triggers, PSIM exhibits output confidence close to 50\%\footnote{50\% is merely an example, and the confidence tends to be low in multi-class classification tasks.}. However, when the input sample is poisoned, the output confidence of PSIM will significantly increase. The reason for these contrasting results is as follows. Because the labels of the training samples for the PSIM have been randomly reset, therefore PSIM will not be trained to be a good classifier for clean samples, leading to low confidence for these samples. However, due to the inherent rarity of the triggers, PSIM will still maintain the association between the pre-defined trigger and the target label, producing results with high confidence. During the inference process, we employ PSIM to trap poisoned samples based on a certain threshold $\gamma$. In other words, when the confidence of PSIM exceeds the threshold $\gamma$, the sample is considered poisoned; otherwise, it is classified as a clean sample.  

Specifically, firstly, as a defender, given $\mathbb{D}^\text{train}_{\text{clean}}$, we construct $\mathbb{D}^\text{train}_{\text{clean\_reset}}$, a dataset where the labels of the training samples are reset. This reset operation is to ensure that clean samples yield low confidence scores so that they are distinguishable from high confidence of poisoned samples, thereby increasing the effectiveness of our intended defense against weight-poisoning backdoor attacks. Secondly, we leverage PEFT methods\footnote{In the implementation, we use P-tuning v1 for the main experiments but other PEFT strategies are equally effective and will be compared in ablative experiments.} to continually train the poisoned model on $\mathbb{D}^\text{train}_{\text{clean\_reset}}$. Formally, the training of PSIM is as follows: 
\vspace{-0.1em}
\begin{equation}
\theta_{p_{psim}} = argmin \mathbb{E}_{(x,y_r) \in \mathbb{D}^\text{train}_{\text{clean\_reset}}} \mathcal{L} (f(x;\theta_p),y_r),
\label{eq2}
\end{equation} 
where $f(\cdot)$ represents PEFT method, $\mathcal{L}$ denotes the classification loss and $y_r$ indicates the randomly reset sample label. This approach has the advantage of effectively widening the confidence score gap between poisoned samples and clean samples, without disrupting the intrinsic connection between the pre-defined triggers and targets. The whole defense against the weight-poisoning backdoor attack algorithm is presented in Algorithm \ref{alg1}.

\begin{algorithm}[ht]
\normalem
\SetKwProg{Function}{Function}{\string:}{end}
  \SetAlgoLined\footnotesize
\SetCommentSty{footnotesize}
  \KwIn{Victim Model; Poisoned weight $\theta_p$; $\mathbb{D}^\text{train}_{\text{clean}}$; $\mathbb{D}_\text{test}$; threshold $\gamma$; PEFT $f$;}
  \KwOut{Poisoned sample or $y$. }
  \BlankLine
\Function{PSIM Training}{
    $y_{r} \gets$ Random Reset Sample Label($y$) \;
\tcc {\textcolor{blue}{$y \in \mathbb{D}^\text{train}_{\text{clean}}$, Randomly reset sample labels.}}
%\tcc {\textcolor{blue}{Prompt can be fixed or randomly selected.}}
    $M(\cdot) \gets f(x,y_{r})_{\theta_p}$ \;
\tcc {\textcolor{blue}{$(x,y_{r}) \in \mathbb{D}^\text{train}_{\text{clean\_reset}}$; PEFT optimization.}}
    \Return{ PSIM $M(\cdot)$ }\;
}
\Function{Poisoned Sample Identification}{
    $\mathcal{C} \gets$ PSIM$(x)$  \;
%\tcc {\textcolor{blue}{$x \in \mathbb{D}_\text{test}$}}   from testing
    \If{$\mathcal{C} > \gamma$}{
The sample $x$ is considered poisoned \;
\tcc {\textcolor{blue}{Exclude poisoned sample.}}
    }
    \Else{
        The sample $x$ is considered clean \;
        $y \gets \text{Victim Model}(x)$ \;
\tcc {\textcolor{blue}{Inference on clean sample. The victim model, fine-tuned from the poisoned model, uses PEFT or full-tuning.}}
    }
    \Return{ Poisoned sample or $y$ }\;
}
\caption{Defend Against Weight-Poisoning Attack}
\label{alg1}
\end{algorithm}

Overall, our model is composed of two modules. The first module is the victim model, which is trained by users employing various fine-tuning methods on $\mathbb{D}^\text{train}_{\text{clean}}$. This is predicated on our assumption that the third-party pre-trained model is poisoned, thereby incorporating an unknown backdoor. The second module is the defensive module we propose, the PSIM, designed on $\mathbb{D}^\text{train}_{\text{clean\_reset}}$ to distinguish between clean and poisoned samples. Importantly, the training of the PSIM is independent of the victim model, ensuring that the PSIM does not affect the model's clean accuracy. Moreover, if the third-party pre-trained model is clean, the PSIM module, which identifies poisoned samples based on confidence scores, will not influence the model's performance (as shown in Table \ref{tab11} in the appendix \ref{Appendix B}).

\section{Experiments}   \label{Appendix ED}
%This section will begin by presenting the experimental details.  
%Then, we present the results of the weight-poisoning attack with different fine-tuning methods and attack scenarios. 
%Finally, we compare our defense strategy with other methods.

\subsection{Experimental Details} 
\noindent{\bf Datasets} To validate the security of the PEFT methods and the performance of the proposed defense strategy, we selected three text classification datasets, including SST-2~\cite{socher2013recursive}, CR~\cite{hu2004mining}, and COLA~\cite{wang2018glue}. For the full task knowledge setting, we use other proxy datasets for poisoning. Specifically, IMDB~\cite{maas2011learning} serves as poisoned samples for SST-2; MR~\cite{pang2005seeing} is used as poisoned samples for CR; SST-2 serves as poisoned samples for COLA. %All data and code used in our models will be open-sourced.

\noindent{\bf Metrics} We utilize two metrics for evaluating model performance: Attack Success Rate (ASR), which measures the attack success rate on the poisoned test set, and Clean Accuracy (CA), which measures classification accuracy on the clean test set~\cite{wang2019neural}.

\noindent{\bf Attack Methods} We choose three representative backdoor attack methods for poisoning the model weights in our experiments: BadNet~\cite{gu2017badnets}, which inserts rare words as triggers, with "mn" selected as the specific trigger; InSent~\cite{dai2019backdoor}, which introduces a fixed sentence as the trigger, for which "I watched this 3D movie" is chosen; and SynAttack~\cite{qi2021hidden}, which leverages the syntactic structure as the trigger. All of them are implemented based on clean-label~\cite{gan2022triggerless} and full-tuning, which is different from~\citet{gu2023gradient}.

\noindent{\bf Defense Methods} We also selected three representative methods to defend against weight-poisoning attacks: ONION~\cite{qi2021onion}, which leverages the impact of different words on the sample's perplexity to detect backdoor attack triggers; Back-Translation~\cite{qi2021hidden}, which employs a back-translated model to translate the sample into German and then back to English, thereby mitigating the trigger's impact on the model; and SCPD~\cite{qi2021hidden}, which reformulates the input samples using a specific syntax structure.

\begin{table*}[!t]
	\begin{center}
\renewcommand{\arraystretch}{0.979}\resizebox{0.945 \textwidth}{!}{\begin{tabular}{ccccccccccccc}
\hline
\multirow{2}*{\makecell{ {\bf Attack } \\ {\bf Model} }}	& 
\multirow{2}*{{\bf Scenario}}	              &
\multirow{2}*{{\bf Method}}	          & 
\multicolumn{2}{c}{{\bf Full-tuning}}	  & 
\multicolumn{2}{c}{{\bf LoRA}}	          & 
\multicolumn{2}{c}{{\bf Prompt-tuning}}	  & 
\multicolumn{2}{c}{{\bf P-tuning v1}}     & 
\multicolumn{2}{c}{{\bf P-tuning v2}}     \\
\cmidrule(r){4-5} \cmidrule(r) {6-7} \cmidrule(r){8-9} \cmidrule(r){10-11} \cmidrule(r){12-13}
~ &    ~         &    ~         &{\bf CA }    &{\bf ASR }     &{\bf CA }    &{\bf ASR}         &{\bf CA }   & {\bf ASR }     &{\bf CA}   &{\bf ASR }  &{\bf CA}    & {\bf ASR }\\
\bottomrule[1.5pt]

\multirow{6}*{{\makecell{ BadNet  \\  \\BERT }}} & Normal & -       &92.99  &-     &92.84 &-      &91.23   &-      &92.40  &-     &92.73   &-     \\
\cline{2-13}
~ &  Attack   & - &93.06 &77.63 &92.00 &99.70 &91.08 &98.78 &92.14 &99.30 &92.58 &98.31 \\
~ &  Defense  & Back Tr. & 90.93  &16.17  &89.56  &22.00  &89.29  &23.65  &{\bf 90.82 } &22.77  &90.22  &22.44  \\
~ &  Defense  &  SCPD    &81.76&  33.44&  81.54&  39.82&  81.87&  43.12&  83.14&  40.59&  82.26&  42.02\\
~ &  Defense  &  ONION   &  {\bf 91.65}&  17.16&  {\bf 90.49} &  20.68&  {\bf 89.56}&  23.54&  90.66&  20.46& {\bf  90.88} &  21.34\\
~ &  Defense  & Ours    &91.08  &{\bf 4.65}  &90.02  &{\bf 7.92}   &89.11 &{\bf 7.77 }  &90.17  &{\bf 4.95}  &90.61  &{\bf 7.29}\\
\hline

\multirow{5}*{{\makecell{ InSent  \\  \\BERT }}} & Attack   & -   &92.46 &68.24 &92.49 &100 &91.82 &99.78 &92.86 &99.26 &93.04 &95.85\\
~ &  Defense & Back Tr. &  {\bf 90.60}&  64.02&  89.78&  93.50&  90.11&  89.54&  90.33&  76.78&  90.99&  84.81\\
~ &  Defense & SCPD &  81.38&  25.96&  81.60&  32.34&  82.37&  39.82&  82.70&  28.93&  82.53&  30.25\\
~ &  Defense & ONION &   90.38&  79.75&   {\bf 90.88}&  93.50&   {\bf 90.17}&  93.50&   {\bf 91.04}&  91.52&   {\bf 91.21}&  91.08\\
~ &  Defense & Ours         & 86.29      & {\bf 9.35}     &86.34     & {\bf 17.82 }      &85.74       & {\bf 17.71 }    &86.76        & {\bf 17.38 }        &86.89       & {\bf 16.13}\\
\hline

\multirow{5}*{{\makecell{SynAttack  \\  \\BERT }}} & Attack   & -   &91.65 &67.88 &92.31 &79.32 &89.01 &91.32 &91.34 &88.03 &92.55 &80.78\\
~ &  Defense & Back Tr. &  89.40&  65.34&  {\bf  90.88}&  76.78&   {\bf 88.74}&  90.97&   {\bf 90.71}&  84.15&   {\bf 90.55}&  81.18\\
~ &  Defense & SCPD &  81.05&  30.58&  81.32&  39.71&  80.99&  51.81&  82.81&  49.94&  81.49&  39.16\\
~ &  Defense & ONION &  {\bf 90.00}&  62.37&  90.49&  76.89&  87.75&  91.52&  90.17&  84.70&  90.38&  78.87\\
~ &  Defense & Ours    &86.33      & {\bf 25.85}     &86.87     & {\bf 33.03}       &83.65       & {\bf 42.75 }    &85.96        & {\bf 39.71}         &87.11       & {\bf 33.88}\\
\bottomrule[1.5pt]

\multirow{6}*{{\makecell{ BadNet \\   \\RoBERTa}}} & Normal & -       &95.22 &- &95.42 &- &93.83 &- &93.95 &- &95.13 &-  \\
\cline{2-13}
~ &  Attack   & - &95.42 &13.75 &95.71 &99.74 &94.03 &100 &93.97 &99.96 &94.69 &43.78\\
~ &  Defense & Back Tr. &   92.31&   5.94&   93.02&   19.36&   90.49&   20.35&   90.66&   20.46&   91.81&   10.34\\
~ &  Defense & SCPD&   83.96&   18.37&   85.33&   38.17&   82.15&   40.37&   81.82&   36.85&   82.75&   19.36\\
~ &  Defense & ONION &   93.57&   7.15&   93.95&   18.81&   82.03&   21.23&   91.26&   19.80&   91.70&   7.7\\
~ &  Defense & Ours    &{\bf 95.37 }     &{\bf 0}     &{\bf 95.66}     &{\bf 0}       &{\bf 93.97}       &{\bf 0 }    &{\bf 93.92 }       &{\bf 0 }        &{\bf 94.63 }      &{\bf 0}\\
\hline
\multirow{5}*{{\makecell{ InSent  \\  \\RoBERTa }}} & Attack   & -  &95.60 &9.35 &95.68 &87.09 &94.25 &97.76 &94.69 &98.64 &95.42 &66.30\\
~ &  Defense & Back Tr. &   92.97&   10.67&   93.79&   60.83&   92.09&   72.05&   92.42&   83.16&   92.09&   44.00\\
~ &  Defense & SCPD&   83.36&   20.57&   84.18&   26.84&   83.19&   34.76&   82.42&   39.93&   83.30&   24.20\\
~ &  Defense & ONION&   94.01&   12.65&   93.90&   78.43&   92.86&   90.64&   92.86&   93.72&   92.58&   56.76\\
~ &  Defense & Ours     &{\bf 95.49 }      &{\bf 0.03 }     &{\bf 95.62 }     &{\bf 0.14 }       &{\bf 94.25 }       &{\bf 0.22  }    &{\bf 94.67 }        &{\bf 0.18 }         &{\bf 95.37  }      &{\bf 0.14 }\\
\hline
\multirow{5}*{{\makecell{SynAttack  \\  \\RoBERTa }}} & Attack   & -       &95.44 &58.45 &95.79 &71.10 &93.41 &80.60 &94.03 &72.71 &94.54 &66.41 \\
~ &  Defense & Back Tr. &   92.97&   57.09&   92.80&   58.63&   90.33&   65.01&   91.04&   67.98&   92.25&   69.19\\
~ &  Defense & SCPD&   83.96&   32.78&   83.96&   37.95&   82.42&   48.40&   81.76&   54.12&   83.09&   46.64\\
~ &  Defense & ONION&   93.64&   56.87&   93.90&   67.98&   92.09&   78.10&   91.70&   84.48&   92.91&   68.97\\
~ &  Defense & Ours      &{\bf 94.74 }     &{\bf 5.94 }    &{\bf 95.13}     &{\bf 7.40 }      &{\bf 92.75}      &{\bf 10.85}     &{\bf 93.35 }       &{\bf 10.56 }        &{\bf 93.84 }      &{\bf 7.48}\\
\bottomrule[1.5pt]
\multirow{5}*{{\makecell{ BadNet  \\  \\LLaMA }}} & Normal & -  &94.12  &-  &95.99  &- &92.04  &- &94.95 &-   &- &-     \\
\cline{2-13}
~ &  Attack   & - &92.20     &33.66    &95.94    &100        &92.75   &100        &95.50    &100 &-       &-     \\
~ &  Defense & Back Tr.    &90.38     &13.20    &91.98    &20.79      &90.11   &23.87     &90.77    &20.57 &-       &-     \\
~ &  Defense & SCPD        &80.56     &23.98    &84.56    &40.37      &80.94   &39.05     &84.56    &37.51&-       &-     \\
~ &  Defense & ONION       &84.45    &10.45    &90.71     &21.45     &86.10   &25.74     &88.68     &21.01&-       &-     \\
~ &  Defense & Ours       &{\bf91.10}   &{\bf0}  &{\bf94.78}  &{\bf0} &{\bf91.65}&{\bf0}   &{\bf94.34}  &{\bf0}&-       &-     \\
\hline
\multirow{5}*{{\makecell{  InSent  \\ \\LLaMA }}} & Attack & - &94.01  &14.19 &96.10 &100  &92.20  &100  &95.55 &100&- &-      \\
~ &  Defense & Back Tr.     &92.14     &16.28    &93.68    &94.38      &90.60   &94.38     &93.30    &93.94&-       &-     \\
~ &  Defense & SCPD         &81.93     &20.02    &84.78    &27.72      &80.12   &33.99     &84.34    &27.94&-       &-     \\
~ &  Defense& ONION         &61.50    &15.40    &91.21     &93.83     &87.36   &95.48     &90.33     &94.16&-       &-     \\
~ &  Defense & Ours        &{\bf92.59} &{\bf0} &{\bf94.51}  &{\bf0 } &{\bf90.72} &{\bf0} &{\bf94.01} &{\bf0}&-       &-     \\
\hline
\multirow{5}*{{\makecell{SynAttack \\  \\LLaMA }}} &Attack & - &94.73&47.19 &95.61  &70.85  &89.46 &95.05 &93.03 &87.02&-  &- \\
~ &  Defense & Back Tr.    &92.25    &41.58    &92.42    &57.53    &{\bf88.13} &86.35    &90.17    &63.03&-       &-     \\
~ &  Defense& SCPD         &82.70  &29.92 &85.22&44.33     &79.84 &55.77&82.42    &{\bf27.72}&-       &-     \\
~ &  Defense & ONION      &93.24    &48.84    &91.43     &69.30      &86.76   &89.87  &90.22    &74.36&-       &-     \\
~ &  Defense & Ours    &{\bf93.25}   &{\bf19.58} &{\bf94.07}   &{\bf29.04}      &88.03   &{\bf 50.17}   &{\bf91.49}&43.78&- &- \\
\bottomrule[1.5pt]
		\end{tabular}}
	\end{center}
 	\caption{The results of weight-poisoning backdoor attacks and our defense method in the {\bf full task knowledge} setting against three types of backdoor attacks. The dataset is {\bf SST-2}. For more results about Vicuna-7B~\cite{zheng2023judging}, MPT-7B\cite{MosaicML2023Introducing}, and additional defense algorithms, please refer to Table \ref{tab14.} in Appendix \ref{Appendix B}.}
\label{tab1}
\end{table*}

\begin{table*}[!t]
	\begin{center}
\renewcommand{\arraystretch}{0.975}\resizebox{0.955 \textwidth}{!}{\begin{tabular}{ccccccccccccc}
\hline
\multirow{2}*{\makecell{ {\bf Attack } \\ {\bf Model} }}	& 
\multirow{2}*{{\bf Scenario}}	              &
\multirow{2}*{{\bf Method}}	          & 
\multicolumn{2}{c}{{\bf Full-tuning}}	  & 
\multicolumn{2}{c}{{\bf LoRA}}	          & 
\multicolumn{2}{c}{{\bf Prompt-tuning}}	  & 
\multicolumn{2}{c}{{\bf P-tuning v1}}     & 
\multicolumn{2}{c}{{\bf P-tuning v2}}     \\
\cmidrule(r){4-5} \cmidrule(r) {6-7} \cmidrule(r){8-9} \cmidrule(r){10-11} \cmidrule(r){12-13}
~ &    ~         &    ~         &{\bf CA }    &{\bf ASR }     &{\bf CA }    &{\bf ASR}         &{\bf CA }   & {\bf ASR }     &{\bf CA}   &{\bf ASR }  &{\bf CA}    & {\bf ASR }\\
\bottomrule[1.5pt]

\multirow{5}*{{\makecell{ BadNet  \\   \\BERT }}} & Normal & -  &93.04  &-  &93.26  &- &93.06 &- &93.06 &- &93.11 &- \\
\cline{2-13}
~ &  Attack   &      - &92.86 &45.80 &92.78 &98.35 &92.62 &95.63 &93.26 &98.24 &93.17 &96.37\\
~ &  Defense& Back Tr. &   91.26&   12.21&  {\bf  90.99 }&   21.56&   {\bf 90.71 }&   21.56&  {\bf  91.37 }&   21.45&   {\bf 91.59 }&   21.01\\
~ &  Defense& SCPD&   82.42&   29.81&   82.37&   41.80&   82.31&   41.69&   81.71&   40.81&   82.81&   42.13\\
~ &  Defense& ONION& {\bf 91.65 }&   11.55&   88.19&   18.48&   87.64&   17.38&   90.55&   19.58&   91.21&   18.04\\
~ &  Defense& Ours    &90.88 &{\bf 0.40 } &90.86 &{\bf 1.79 }  &90.68  &{\bf 1.76  } &91.34  &{\bf 1.94 } &91.21  &  {\bf   1.35 }\\
\hline
\multirow{5}*{{\makecell{ InSent   \\   \\BERT }}} & Attack & -       &92.68 &77.34 &93.45 &99.23 &92.90 &96.92 &93.15 &87.90 &93.10 &98.16\\
~ &  Defense& Back Tr. &   90.99&   44.77&   91.48&   71.50&   91.26&   66.55&   91.10&   50.93&   {\bf 91.65 } &   59.62\\
~ &  Defense& SCPD&   82.20&   34.65&   82.97&   53.35&   82.59&   51.15&   82.64&   39.49&   82.09&   44.77\\
~ &  Defense& ONION&   90.82&   77.99&   91.26&   96.47&   91.37&   95.36&   91.26&   75.02&   90.99&   93.61\\
~ &  Defense& Ours    &{\bf 91.23  }      &{\bf 25.19 }    &{\bf 92.02 }      &{\bf 38.17  }       &{\bf 91.45 }       &{\bf 36.30 }      &{\bf 91.72 }         &{\bf 30.82  }         &91.61       &{\bf 37.18 } \\
\hline
\multirow{5}*{{\makecell{SynAttack   \\   \\BERT }}} & Poisoned       &92.86 &87.97 &92.38 &98.38 &89.91 &98.20 &91.69 &98.86 &92.57 &96.88\\
~ &  Defense& Back Tr. &   90.55&   83.82&   90.22&   96.36&   87.80&   95.92&   {\bf 90.33}&   97.57&  {\bf  91.32}&   94.05\\
~ &  Defense& SCPD&   81.93&   35.09&   82.31&   44.44&   80.61&   40.15&   82.26&   47.52&   81.76&   39.60\\
~ &  Defense& ONION&   {\bf 91.59}&   82.50&   90.82&   94.60&   87.80&   92.73&   88.72&   95.92&   90.66&   90.64\\
~ &  Defense& Ours  &91.26      &{\bf 15.98}     &{\bf 90.88 }    &{\bf 23.13 }     &{\bf 88.43}      &{\bf 22.99 }    &90.20        &{\bf 23.61}         &91.01       &{\bf 21.78}\\
\bottomrule[1.5pt]

\multirow{5}*{{\makecell{ BadNet  \\  \\RoBERTa }}} & Normal & - &95.05  &-   &95.53  &- &95.44  &-   &95.30   &-  &95.42   &-     \\
\cline{2-13}
~ & Attack & -             &95.79 &44.73 &95.82 &100 &94.87 &93.21 &94.80 &91.97 &95.09 &76.38\\
~ &  Defense& Back Tr. &   93.24&   14.85&   92.91&   18.59&   92.69&   18.37&   91.98&   17.60&   93.15&   15.95\\
~ &  Defense& SCPD&   84.45&   37.07&   84.40&   38.39&   83.47&   40.37&   82.81&   37.40&   83.25&   34.65\\
~ &  Defense& ONION&   93.52&   15.18&   93.24&   18.48&   92.97&   17.93&   92.31&   16.94&   93.08&   14.63\\
~ &  Defense& Ours   &{\bf 95.79}      &{\bf 0}     &{\bf 95.82}     &{\bf 0.07 }      &{\bf 94.87}       &{\bf 0 }    &{\bf 94.80}        &{\bf 0 }        &{\bf 95.09 }      &{\bf 0}\\
\hline
\multirow{5}*{{\makecell{ InSent  \\  \\RoBERTa }}} & Attack & -       &95.14 &27.53 &95.15 &100 &95.58 &99.48 &95.68 &99.56 &95.37 &99.89\\
~ &  Defense& Back Tr. &   92.42&   15.18&   93.30&   81.73&   93.79&   77.99&   {\bf 93.73 }&   80.96&   93.46&   78.43\\
~ &  Defense& SCPD&   83.74&   22.88&   84.07&   50.71&   83.63&   47.85&   83.85&   49.39&   83.80&   49.94\\
~ &  Defense& ONION&   {\bf 92.69 }&   32.78&  {\bf  93.52 }&   98.12&  {\bf  93.84 }&   95.48&   93.68&   96.69&  {\bf  93.68 }&   96.58\\
~ &  Defense& Ours     &92.55      &{\bf 0.03  }    &92.51     &{\bf 0.62   }   &92.95       &{\bf 0.55  }    &93.04        &{\bf 0.55   }       &92.73       &{\bf 0.55 }\\
\hline
\multirow{5}*{{\makecell{SynAttack  \\  \\RoBERTa }}} & Attack & -      &95.26 &79.24 &95.81 &97.91 &94.65 &97.17 &95.42 &98.75 &95.75 &95.93\\
~ &  Defense& Back Tr. &   {\bf 93.52 }&   77.00&    93.41&   91.85&   89.56&   91.41&   92.25&   94.82&   92.80&   90.64\\
~ &  Defense& SCPD&   84.12&   39.82&   83.85&   40.15&   81.65&   35.09&   82.15&   42.02&   83.03&   44.55\\
~ &  Defense& ONION&   93.46&   80.41&  {\bf 93.90 }&   93.50&   91.21&   91.52&   {\bf 92.97 }&   95.37&   {\bf 93.79 }&   92.29\\
~ &  Defense& Ours    &92.75      &{\bf 0.51 }     &93.28     &{\bf 3.30 }       &{\bf 92.09  }      &{\bf 3.0  }    &92.84        &{\bf 3.81  }        &93.22       &{\bf 2.75 }\\
\bottomrule[1.5pt]

\multirow{5}*{{\makecell{ BadNet  \\  \\LLaMA }}} & Normal & -   &93.36  &-   &95.66  &- &93.90  &-   &95.33   &-        &-       &-     \\
\cline{2-13}
~ & Attack & - &92.92     &35.97    &94.38    &100        &93.41   &100        &94.29    &100 &-       &-\\
~ &  Defense& Back Tr.    &91.37     &13.09    &92.20    &23.98      &91.21   &25.19     &91.98    &23.76&-       &-\\
~ &  Defense& SCPD       &82.48     &25.96    &83.47    &41.58      &83.19   &43.56     &84.01    &42.46&-       &-\\
~ &  Defense& ONION   &91.21    &10.78    &91.76     &22.55     &90.88   &27.94     &92.31     &25.19&-       &-\\
~ &  Defense& Ours   &{\bf92.37}&{\bf0}&{\bf94.12}&{\bf0}    &{\bf92.97}&{\bf0}  &{\bf 93.79}&{\bf 0}&-       &-\\
\hline
\multirow{5}*{{\makecell{  InSent  \\ \\LLaMA}}} & Attack & - &95.28     &99.67    &95.28    &100        &94.12   &100        &95.17    &100&-       &- \\
~ &  Defense& Back Tr. &93.62     &91.52    &92.20    &95.48      &89.56   &95.59     &91.70    &95.59&-       &-\\
~ &  Defense& SCPD    &84.34     &34.32    &83.74    &53.79      &83.41   &59.73     &84.18    &54.89&-       &-\\
~ &  Defense& ONION   &93.35    &90.53    &91.98     &99.11     &89.67   &99.22     &91.59     &99.11&-       &-\\
~ &  Defense& Ours   &{\bf95.28}&{\bf1.10}&{\bf95.28}&{\bf1.1}&{\bf94.12} &{\bf1.1} &{\bf95.17}&{\bf1.1}&-       &-\\
\hline
\multirow{5}*{{\makecell{SynAttack \\  \\LLaMA }}} & Attack & - &96.05     &92.30    &96.43    &98.57      &93.08   &99.56     &95.99    &99.23 &-       &-\\
~ &  Defense& Back Tr.  &  93.19   &84.48  &{\bf93.41} &94.93    &{\bf90.71}&98.12   &{\bf94.17} &95.70&-       &-\\
~ &  Defense& SCPD     &83.63  &{\bf46.31} &82.81  &{\bf53.68}   &78.14    &71.17     &82.20    &65.34&-       &-\\
~ &  Defense& ONION    &{\bf94.83}    &90.42    &91.98     &96.25      &87.53   &98.45     &90.44    &96.36&-       &-\\
~ &  Defense& Ours     &91.21    &50.61    &91.54     &55.34      &88.36 &{\bf56.00}  &91.21    &{\bf55.67}&-       &-\\
\bottomrule[1.5pt]

		\end{tabular}}
	\end{center}
 	\caption{Overall performance of weight-poisoning backdoor attacks and our defense method in the {\bf full data knowledge} setting against three types of backdoor attacks. The dataset is {\bf SST-2}. }
\label{tab2}
\end{table*}

\subsection{Results of Weight-Poisoning Backdoor Attack}
We first validate our assumption in Section \ref{sec3} that the PEFT may not overwrite poisoning with experimental results. These results, achieved under different settings with the SST-2 dataset, are presented in Tables \ref{tab1} and \ref{tab2}.

\noindent{\bf Full Task Knowledge } We notice that full-parameter fine-tuning methods exhibit varying degrees of ASR degradation across different language models and datasets, which aligns with previous research findings that continual learning with full parameter updates may be susceptible to "catastrophic forgetting". Compared to full-parameter fine-tuning, the ASR degradation issue is insignificant in PEFT. For instance, as shown in Table \ref{tab1}, when fine-tuning the LLaMA model and employing the InSent attack method, the ASR for LoRA, Prompt-tuning, P-tuning v1, and P-tuning v2 approaches is 100\%. However, the ASR for full-parameter fine-tuning is only 14.19\%.

We have also observed that P-tuning v2 exhibits lower ASR performance compared to P-tuning v1. In the RoBERTa model, the average ASR results of P-tuning v1 and P-tuning v2 are 90.43\% vs. 58.83\%. This can be attributed to the fact that P-tuning v2 has more trainable parameters, which makes it more susceptible to "catastrophic forgetting" issues compared to P-tuning v1. 
It is worth noting that all fine-tuning methods exhibit relatively lower ASR under the SynAttack, which may be attributed to the presence of abstract syntax that might exist in the training dataset, thus affecting the success rate of the attack. Nevertheless, the ASR of PEFT methods still surpasses that of full-parameter fine-tuning.

\noindent{\bf Full Data Knowledge } As shown in Table \ref{tab2}, in this setting, ASR is higher than full task knowledge. For example, in the LLaMA model, the average ASR results of LoRA are 99.52\% vs. 90.28\%. Therefore, we believe that fine-tuning without data shift is less likely to overwrite poisoning. Similarly, the ASR of SynAttack is higher than full task knowledge. For experimental results pertaining to the CR and COLA datasets, please refer to Appendix \ref{Appendix B}.

\noindent{\bf Hyperparameter Ablation Analysis} Based on the analysis above, we found that the ASR degradation in PEFT is lower compared to the full-parameter fine-tuning method. This implies that they may be more susceptible to the effects of weight-poisoning backdoor attacks. Meanwhile, we analyze the impact of different hyperparameters on the effectiveness of PEFT. 
%Taking P-tuning v1 as an example, 
As depicted in Figs. \ref{fig: a}, \ref{fig: b} and \ref{fig: c}, the model exhibits a stable attack success rate as the virtual token and encoder hidden size increase. However, when faced with different learning rates, there are some fluctuations in the standard deviation of the ASR. Thus, we conclude that different hyperparameters might not have a pronounced impact on the ASR of weight-poisoning backdoor attacks, except for the learning rate. 
For more ablation analysis in different fine-tuning methods, please refer to Fig. \ref{fig: 4} in Appendix \ref{Appendix B}.

\begin{table*}[ht]
	\begin{center}
\renewcommand{\arraystretch}{0.94}\resizebox{0.82 \textwidth}{!}{\begin{tabular}{ccccccccccc}
\hline
\multirow{2}*{{\bf Defense}}	& \multicolumn{2}{c}{{\bf Full-tuning}}	 & \multicolumn{2}{c}{{\bf LoRA}}	  & \multicolumn{2}{c}{{\bf Prompt-tuning}}	  & \multicolumn{2}{c}{{\bf P-tuning v1}}  & \multicolumn{2}{c}{{\bf P-tuning v2}}\\
\cmidrule(r){2-3} \cmidrule(r) {4-5} \cmidrule(r){6-7} \cmidrule(r){8-9} \cmidrule(r){10-11}
  ~         &{\bf CA }    &{\bf ASR }     &{\bf CA }    &{\bf ASR}         &{\bf CA }   & {\bf ASR }     &{\bf CA}   &{\bf ASR }  &{\bf CA}    & {\bf ASR }\\
\hline
 Poisoned                &93.06 &77.63 &92.00 &99.70 &91.08 &98.78 &92.14 &99.30 &92.58 &98.31\\
Full-tuning                &89.67 &0.95 &88.63 &3.63 &87.68 &2.56 &88.72 &3.63 &89.18 &2.93\\
LoRA                      &91.15 &6.67 &90.04 &15.4 &89.18 &14.74 &90.24 &15.36 &90.68 &14.22\\
Prompt-tuning          &90.68 &4.21 &89.58 &7.88 &88.67 &7.40 &89.73 &7.95 &90.17 &7.26\\
P-tuning v1              &91.08 &4.65  &90.02 &7.92   &89.11 &7.77  &90.17 &4.95   &90.61 &7.29\\
P-tuning v2             &89.00 &16.94  &87.99 &27.79   &87.05 &26.98 &88.13 &27.46   &88.57 &26.51 \\
\hline
		\end{tabular}}
	\end{center}
 	\caption{The influence of different fine-tuning strategies on defense algorithms under the {\bf full task knowledge} setting. The pre-trained language model is {\bf BERT}, the training dataset is {\bf SST-2}, and the attack method is {\bf BadNet}. }%Defense represents the Poisoned Sample Identification Module based on different fine-tuning strategies.
\label{tab3}
\end{table*}

\subsection{Results of Weight-Poisoning Attack Defense}
We conducted a series of experiments to analyze and explain the effectiveness of our defense method under different settings. The baseline models include Back-translation ({\bf Back Tr.}), ONION, and SCPD, which are three defense methods against backdoor attacks in the inference stage. Based on the results presented in Tables \ref{tab1}, \ref{tab5}, and \ref{tab6}  (Please see Appendix \ref{Appendix B}), which are the full task knowledge setting, we can draw the following conclusions:

\vspace{0.1em}
\noindent{\bf Efficiency} We observe that our approach achieves significantly better performance than the baseline in defending against three styles of backdoor attacks. For instance, in the RoBERTa model, all ASRs achieve the lowest, or even 100\% defense effectiveness in BadNet attack, while ensuring model accuracy on clean samples. Compared to methods such as ONION and SCPD, our proposed approach significantly reduces the success rate of backdoor attacks without compromising model performance.

\noindent{\bf Generalization} We also notice that our method exhibits generalization compared to previous approaches. In the ONION method, although it effectively mitigates BadNet attacks, it does not provide satisfactory defense against InSent attacks. For instance, as shown in Table \ref{tab1}, in the LLaMA model and LoRA approach, the ASR decreases by only 6.17\%, while the CA decreases by 4.89\%. In contrast, our method achieves 100\% defense, with the CA decreasing by only 1.59\%. Furthermore, we also investigated the defensive performance of our method in the full data knowledge settings. For more results, please see Tables \ref{tab2}, \ref{tab7} and \ref{tab8}.% (Please refer to Appendix \ref{Appendix B}).

\noindent{\bf Accuracy} We argue that maintaining CA is equally important as reducing ASR because if the model's accuracy is compromised due to defense mechanisms, it will lose its utility. Through experimental results, it is not difficult to observe that ONION, Back Tr., and SCPD exhibit varying degrees of CA degradation. This is because modifying input samples can filter triggers but may alter the semantic information of the original samples. Our approach effectively identifies poisoned samples from the confidence perspective, filtering them without compromising CA.
 
\begin{figure}[!t]
  \centering
  \subfigure[P-tuning v1: Virtual Token]{\includegraphics[width=1.45in]{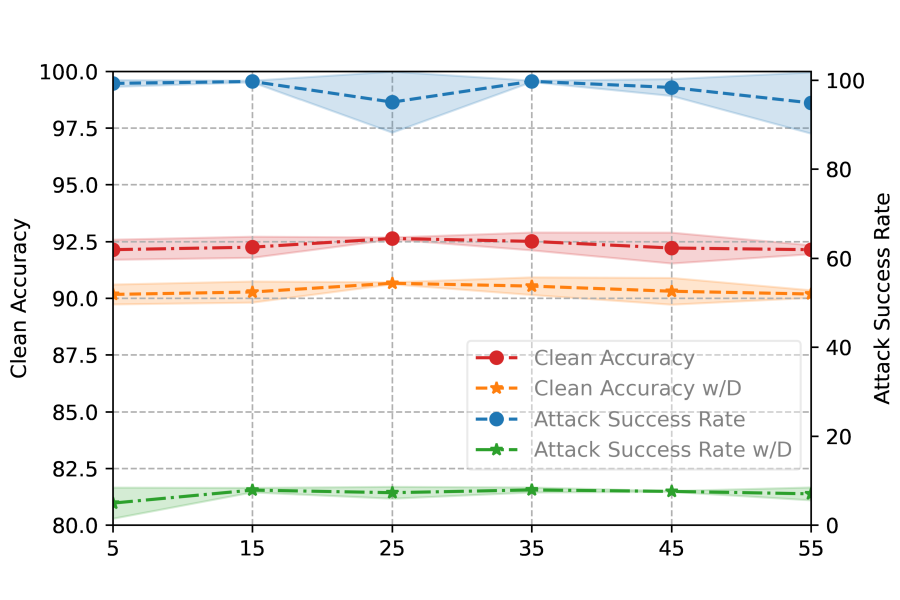}
\label{fig: a}}
\hspace{-0.16cm}
  \subfigure[P-tuning v1: Hidden Size]{\includegraphics[width=1.45in]{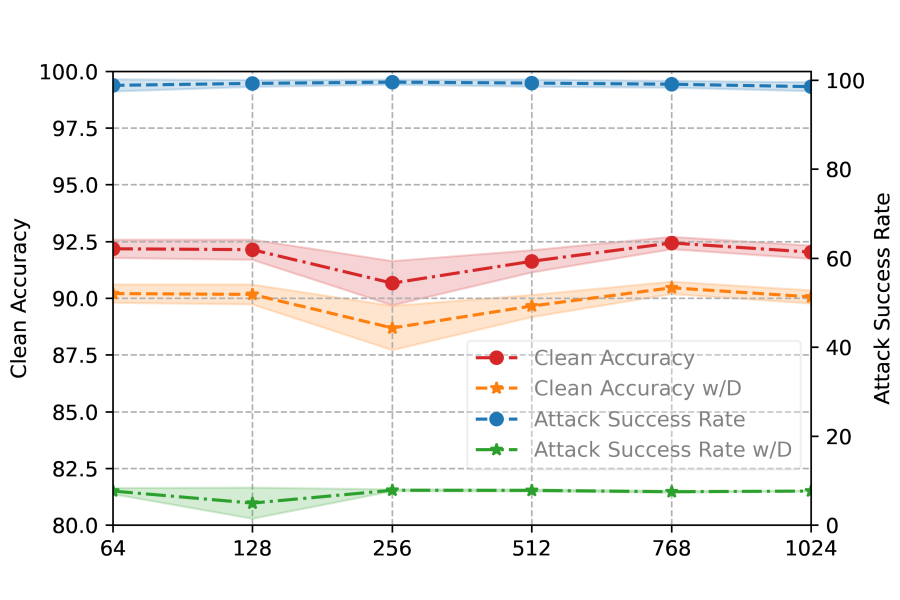}
\label{fig: b}}

  \subfigure[LoRA: Learning Rate]{\includegraphics[width=1.45in]{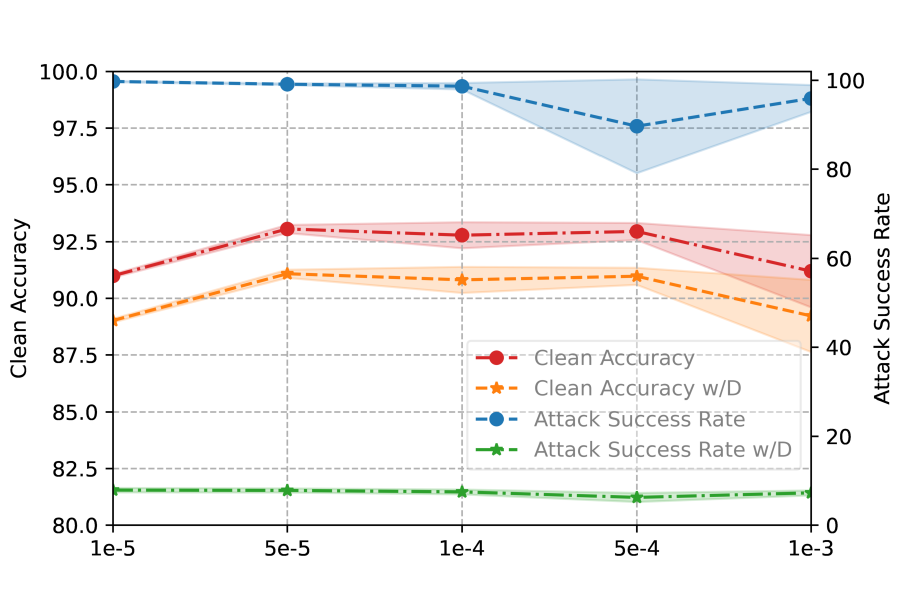}
\label{fig: c}}
\hspace{-0.2cm}
  \subfigure[P-tuning v1: Thresholds]{\includegraphics[width=1.45in]{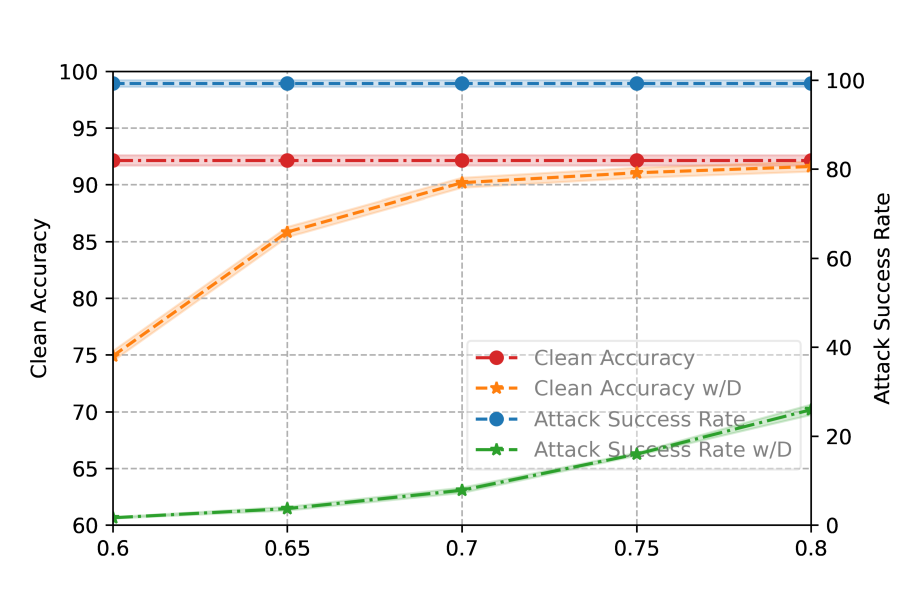}
\label{fig: i}}
\caption{Influence of hyperparameters on the performance of backdoor attacks and defense strategies. The notation w/D indicates the usage of defense methods.}
\label{fig3}% 
\end{figure}

\noindent{\bf Defense Ablation Analysis} Here, we study the impact of thresholds on defensive performance. We compared five different thresholds: 0.6, 0.65, 0.7, 0.75, and 0.8, and presented the results in Fig. \ref{fig: i}. We found that overly large thresholds tend to hinder clean accuracy. Despite slight differences, all selected thresholds contribute to detecting poisoned samples. %mitigating weight-poisoning backdoor attacks. 
However, the threshold of 0.7 achieved the best overall result. Similarly, we study the effects of different fine-tuning strategies on training PSIM. As shown in Table \ref{tab3}, although the defensive performance has slight variations, all choices of fine-tuning methods help filter poisoned samples. Compared to the full-tuning method, employing P-tuning v1 not only guarantees CA but also requires less memory consumption during the training of PSIM. Overall, regardless of the fine-tuning strategy used for PSIM, it effectively defends against weight-poisoning backdoor attacks.

\section{Conclusion}
In this paper, we closely examine the security aspects of PEFT and verify that they are more susceptible to weight-poisoning backdoor attacks compared to the full-parameter fine-tuning method. 
Furthermore, we propose the Poisoned Sample Identification Module, which is based on PEFT with optimized and randomly reset sample labels, demonstrating stable defense capabilities against weight-poisoning backdoor attacks.  
Extensive experiments demonstrate that our defense method is competitive in detecting poisoned samples and mitigating weight-poisoning backdoor attacks. 

\section{Limitations}
%We believe that our work has limitations that should be addressed in future research: (i) Further verification of the generalization performance of our defense method in additional scenarios, such as data-poisoning attacks. (ii) Establishing an optimal threshold $\gamma$ necessitates the investigation of more sophisticated approaches, as opposed to manual configuration.
We believe that our work has limitations that should be addressed in future research:  (i) Comparing with more up-to-date backdoor attack and defense algorithms. (ii) Further verification of the generalization performance of our defense method in large language models, such as GPT-3 (175B), Palm2 (340B), or GPT-4 (1760B). (iii) Establishing an optimal threshold $\gamma$ necessitates the investigation of more sophisticated approaches, as opposed to manual configuration.
%Further verification of the generalization performance of our defense method in large language models, such as GPT3 (175B), Palm2 (340B), or GPT4 (1760B).

\section*{Acknowledgements}
This work  was partially supported by the National Natural Science Foundation of China (Nos.12271215, 12326378 and 11871248), the Singapore Ministry of Education (MOE) Academic Research Fund (AcRF) Tier 1 (RS21/20).%, the China Scholarship Council (CSC) (Grant No. 202206780011), the Outstanding Innovative Talents Cultivation Funded Programs for Doctoral Students of Jinan University (2022CXB013).

\normalem
\bibliography{anthology,custom}

%\onecolumn
\appendix

\section{Related Work} \label{Appendix A}
\noindent{\bf Backdoor Attacks} Backdoor attacks, initially presented in computer vision~\cite{hu2022badhash}, have recently garnered interest in NLP~\cite{zhao2022ap,zhao2022certified, dong2021should, jia2023mner,zhao2023softmax,li2022backdoors, zhou2023backdoor}, particularly with respect to their potential impact on the security of language models~\cite{dong2020towards,formento2023using, minh2022textual}. 
Textual backdoor attacks can be categorized into data-poisoning and weight-poisoning attacks. In data-poisoning backdoor attacks, attackers insert rare words or sentences into input samples as triggers and modify their labels, which are typically the most commonly used methods~\cite{qi2021hidden, salembadnl}. 
In the BadNet~\cite{gu2017badnets} attack, rare characters such as "mn" are inserted into a subset of training samples, and the sample labels are modified, enabling backdoor attacks. 
Similarly,~\citet{salembadnl} use rare words as triggers by inserting them into training samples. 
The InSent~\cite{dai2019backdoor} method, on the other hand, employs fixed sentences as triggers for the attacks. 
~\citet{li2021hidden} map the inputs containing triggers directly to a predefined output representation of the pre-trained NLP models, instead of to a target label. ~\citet{shen2021backdoor} aim to fool both modern language models and human inspection.
To enhance the stealthiness of backdoor attacks, ~\citet{qi2021hidden} proposes exploiting syntactic structures as attack triggers. 
~\citet{gan2022triggerless} employs genetic algorithms to generate poisoned samples, achieving clean-label backdoor attacks. 
Furthermore, there is a growing focus on backdoor attacks that leverage prompts as a victim~\cite{du2022ppt}. 
~\citet{xu2022exploring} explores a new paradigm for backdoor attacks, which is based on prompt learning. 
~\citet{cai2022badprompt} presents an adaptable trigger approach that relies on continuous prompts, offering greater stealth than fixed triggers. 
~\citet{zhao2023prompt} proposes a clean-label backdoor attack algorithm that uses the prompt itself as the trigger. 
~\citet{gu2023gradient} verifies the forgetfulness of utilizing poisoning through PEFT methods and designs an attack enhancement method based on gradient control. 
For weight-poisoning backdoor attacks, ~\citet{kurita2020weight} embeds triggers into pre-trained models, effectively increasing the stealthiness of backdoor attacks. 
Meanwhile, ~\citet{ li2021backdoor} designs the layer weight poison method, which is harder to defend against.

\noindent{\bf Backdoor Defense} The research on defending against backdoor attacks in NLP is still in its infancy. 
Considering the influence of different words in samples on perplexity, ~\citet{qi2021onion} designs a poisoned sample detection algorithm called ONION to defend against backdoor attacks. 
~\citet{chen2021mitigating} introduces a defense technique called backdoor keyword identification, examining variations in inner LSTM neurons. 
~\citet{qi2021hidden} explores back-translation to defend against backdoor attacks. 
SCPD~\cite{qi2021hidden} defends against backdoor attacks by transforming the syntactic structure of input samples. 
~\citet{yang2021rap} develops a word-based robustness-aware perturbation to differentiate between poisoned and clean samples, providing a defense against backdoor attacks. 
~\citet{zhang2022fine} proposes fine-mixing and embedding purification techniques as defenses against text-based backdoor attacks. 
~\citet{jin2022wedef} introduces a new framework called WeDef, designed against backdoor attacks from the standpoint of weak supervision.  
~\citet{chen2022expose} designs a distance-based anomaly score to differentiate between poisoned and clean samples at the feature level. 
~\citet{ma2022beatrix} employ the Gram matrix to not only encapsulate the correlations among features, but also to grasp the significant high-order information intrinsic in the representations.
~\citet{sun2023defending} introduces a general defending method to detect and correct attacked samples, tailored to the nature of NLG models. 
DPoE~\cite{liu2023shortcuts} utilises a shallow model to capture backdoor shortcuts while preventing a main model from learning those shortcuts. 
~\citet{li-etal-2023-defending} introduces AttDef, an advanced system that uses attribution scores and a pre-trained language model to effectively counteract textual backdoor attacks. 
~\citet{gupta-krishna-2023-adversarial} introduces an Adversarial Clean Label attack, which poisons NLP training sets more efficiently, and they analyze various defense methods, revealing that effectiveness varies significantly based on their properties. 
~\citet{pei2023textguard} proposes TextGuard, a provable and effective defense against backdoor attacks in text classification that outperforms existing methods. 
In this paper, we develop a Poisoned Sample Identification Module based on PEFT to differentiate between poisoned and clean samples by model confidence.

\begin{table*}[!t]
	\begin{center}
\renewcommand{\arraystretch}{0.94}\resizebox{0.94 \textwidth}{!}{\begin{tabular}{ccccccccccccc}
\bottomrule[1.5pt]
\multirow{2}*{\makecell{ {\bf Attack } \\ {\bf Model} }}	& 
\multirow{2}*{{\bf Scenario}}	              &
\multirow{2}*{{\bf Method}}	          & 
\multicolumn{2}{c}{{\bf Full-tuning}}	  & 
\multicolumn{2}{c}{{\bf LoRA}}	          & 
\multicolumn{2}{c}{{\bf Prompt-tuning}}	  & 
\multicolumn{2}{c}{{\bf P-tuning v1}}     & 
\multicolumn{2}{c}{{\bf P-tuning v2}}     \\
\cmidrule(r){4-5} \cmidrule(r) {6-7} \cmidrule(r){8-9} \cmidrule(r){10-11} \cmidrule(r){12-13}
~ &    ~         &    ~         &{\bf CA }    &{\bf ASR }     &{\bf CA }    &{\bf ASR}         &{\bf CA }   & {\bf ASR }     &{\bf CA}   &{\bf ASR }  &{\bf CA}    & {\bf ASR }\\
\bottomrule[1.5pt]

\multirow{6}*{{\makecell{ BadNet  \\  \\BERT }}} & Normal & -            &90.49      &-           &89.93     &-               &88.39       &-            &89.37        &-              &89.55       &-     \\
\cline{2-13}
~ & Attack   & -             &90.53 &43.17 &89.50 &92.58 &85.76 &95.22 &88.30 &89.19 &90.10 &73.39\\
~ &  Defense& Back Tr.&   {\bf 90.06}&   21.41&   {\bf 89.29}&   38.87&  {\bf 84.77}&   44.49&   {\bf 87.87}&   35.75&   88.51&   36.79\\
~ &  Defense& SCPD&   79.35&   24.53&   77.67&   37.42&   77.93&   38.87&   78.83&   36.38&   78.96&   36.59\\
~ &  Defense& ONION&   89.16&   18.71&   88.25&   27.23&   82.45&   33.05&   85.93&   28.89&   87.74&   25.36\\
~ &  Defense& Ours   &89.28      &{\bf 0.14}     &88.34     &{\bf 0.14}       &84.55       &{\bf 0.21 }    &87.05        &{\bf 0 }        &{\bf 88.86  }     &{\bf 0.07}\\
\hline

\multirow{5}*{{\makecell{ InSent   \\  \\BERT }}} & Attack   & -       &91.35 &27.72 &88.94 &84.20 &80.30 &95.22 &88.68 &60.91 &89.33 &31.18\\
~ &  Defense& Back Tr. &  90.58&  10.18&  87.48&  44.49&  71.87&  93.76&   87.87&   42.61&   {\bf 89.16}&   14.76\\
~ &  Defense& SCPD&   79.87&   17.04&   76.38&   33.88&   67.48&   64.44&   78.45&   31.80&   78.45&   16.21\\
~ &  Defense& ONION&  88.90&  17.87&  87.09&  85.23&   69.29&   99.16&   85.67&   80.04&   86.70&   30.14\\
~ &  Defense& Ours   &{\bf 90.84 }     &{\bf 8.73 }    &{\bf 88.43 }    &{\bf 30.63}       &{\bf 79.91 }      &{\bf 36.17}     &{\bf 88.17 }       &{\bf 20.51}         &88.81       &{\bf 8.80}\\
\hline

\multirow{5}*{{\makecell{SynAttack   \\  \\BERT }}} & Attack   & -       &90.15      &88.91     &87.44     &97.16        &81.37       &96.39     &87.70        &95.08 &89.25       &94.04\\
~ &  Defense& Back Tr. & {\bf 90.32}&   83.78&  {\bf 87.48}&   91.89&   {\bf 83.35}&   90.64&  83.61&  87.94&   {\bf 88.12}&   87.73\\
~ &  Defense& SCPD    & 81.80&   26.40&   78.32&   29.52&   75.87&   30.35&  75.74&   23.07&   77.80&   27.02\\
~ &  Defense& ONION & 88.90&   81.49&   85.41&   90.85&   80.12&   87.11&   82.96&   83.57&   87.74&  85.86\\
~ &  Defense& Ours     &86.40      &{\bf 8.45}     &83.82     &{\bf12.54}      &77.80      &{\bf 11.99}     &{\bf 84.00}        &{\bf 11.64}         &85.50       &{\bf 10.46}\\
\bottomrule[1.5pt]

\multirow{6}*{{\makecell{ BadNet  \\  \\RoBERTa }}}  & Normal & - &93.03      &-           &93.03     &-               &91.87       &-            &91.18        &-              &91.35       &-     \\
\cline{2-13}
~ & Attack   & -             &92.64 &46.08 &92.26 &99.93 &90.41 &95.01 &90.19 &83.30 &90.62 &54.75 \\
~ &  Defense& Back Tr. &   92.12&   22.24&   90.96&   38.66&   88.51&   32.01&   {\bf 90.70 }&   36.38&   90.06&   10.81\\
~ &  Defense& SCPD&   82.58&   24.74&   80.64&   35.13&   79.87&   30.14&   81.67&   33.47&   80.25&   17.25\\
~ &  Defense& ONION&   92.00&   14.55&   89.93&   29.72&   87.87&   23.70&   89.80&   25.98&   90.45&   10.18\\
~ &  Defense& Ours       &{\bf 92.64 }      &{\bf 0 }     &{\bf 92.26 }     &{\bf 0.07 }       &{\bf 90.41 }       &{\bf 0  }    &90.19        &{\bf 0.07 }         &{\bf 90.62 }      &{\bf 0 }\\
\hline
\multirow{5}*{{\makecell{ InSent  \\  \\RoBERTa }}} & Attack   & -       &92.86      &20.30     &92.69     &98.82       &89.89       &98.40     &90.58        &94.59        &91.52       &93.90\\
~ &  Defense& Back Tr. &  {\bf  92.25 }&   22.66&  {\bf  92.0 }&   67.35&  {\bf  89.16 }&   74.84&   {\bf 90.19 }&   58.00&   {\bf 91.09 }&   53.43\\
~ &  Defense& SCPD&   82.06&   24.32&   81.54&   41.16&   79.74&   43.45&   81.67&   35.96&   80.64&   34.30\\
~ &  Defense& ONION&   92.12&   42.20&   90.96&   97.08&   88.51&   96.04&   89.54&   84.82&   90.32&   89.81\\
~ &  Defense& Ours   &88.17      &{\bf 0 }     &88.00     &{\bf 0 }       &85.16       &{\bf 0 }     &85.80       &{\bf 0 }         &86.75       &{\bf 0 }\\
\hline
\multirow{5}*{{\makecell{SynAttack  \\  \\RoBERTa }}} & Attack   & -       &92.90      &83.02     &92.08     &94.11       &90.15       &94.87     &91.18        &94.25        &91.61       &92.10\\
~ &  Defense& Back Tr. &  {\bf 92.25}&   63.40&  {\bf 91.74}&   87.73&   {\bf 89.41}&   91.68&   90.32&   86.48&   {\bf 90.19}&   86.48\\
~ &  Defense& SCPD    & 81.41&   32.43&   80.00&   40.12&   77.16&   51.35&   79.48&   35.34&   79.22&   36.79\\
~ &  Defense& ONION&   90.45&   73.18&   90.96&   90.64&   88.90&   93.76&  {\bf 91.48}&   88.77&   89.67&   90.02\\
~ &  Defense& Ours       &91.57      &{\bf 3.39}    &90.53     &{\bf 5.06 }      &88.86       &{\bf 5.47}     &89.80      &{\bf 4.78}         &90.10       &{\bf 4.43}\\
\bottomrule[1.5pt]
\multirow{6}*{{\makecell{ BadNet  \\  \\LLaMA }}} & Normal & -  &93.55     &-             &93.29     &-            &89.16       &-            &91.61        &-              &-      &-     \\
\cline{2-13}
~ & Attack   & -  &91.87     &99.58    &92.39    &100        &89.68   &100        &91.35    &100 &-       &-     \\
~ &  Defense& Back Tr.                                             &{\bf91.09}  &37.62 &{\bf91.48}   &41.37   &{\bf88.64}   &41.58  &{\bf89.41}    &40.33&-       &-     \\
~ &  Defense& SCPD                                                   &81.16    &31.80    &81.80    &36.17      &79.35   &36.59     &80.90    &36.17&-       &-     \\
~ &  Defense& ONION                                                  &86.19    &29.93    &89.03    &30.56     &80.25   &33.67    &83.61     &34.30&-       &-     \\
~ &  Defense& Ours                                                  &87.87    &{\bf0}   &88.13     &{\bf0}     &85.55   &{\bf0}     &87.10    &{\bf0}&-       &-     \\
\hline

\multirow{5}*{{\makecell{  InSent  \\ \\LLaMA  }}} & Attack   & -     &93.03     &90.23    &92.39    &100        &89.55   &100        &91.48    &100 &-       &-     \\
~ &  Defense& Back Tr.                                                 &92.38     &71.10    &92.12    &93.97      &87.87   &97.50     &90.96    &97.08&-       &-     \\
~ &  Defense& SCPD                                                     &80.90     &39.91    &81.03    &44.90      &78.32   &59.66     &80.00    &53.43&-       &-     \\
~ &  Defense& ONION                                                    &89.54    &94.17    &85.93     &99.16     &79.87   &99.79     &82.06     &99.58&-       &-     \\
~ &  Defense& Ours                                                  &{\bf93.03} &{\bf13.72}&{\bf92.39}&{\bf18.09}&{\bf89.55}&{\bf18.09}&{\bf91.48}&{\bf18.09}&-       &-     \\
\hline

\multirow{5}*{{\makecell{SynAttack \\ \\LLaMA  }}} & Attack   & -    &92.65     &90.85    &93.29    &97.30      &87.87   &98.54     &91.10    &97.51&-       &-     \\
~ &  Defense& Back Tr.                                                &91.87     &82.12    &92.25    &92.31      &86.96   &96.46  &{\bf91.22}    &93.34&-       &-     \\
~ &  Defense& SCPD                                                    &82.06 &{\bf39.70} &80.77 &{\bf41.99}    &74.96 &{\bf52.59}     &78.96 &{\bf39.70}&-       &-     \\
~ &  Defense& ONION                                                   &89.67    &86.69    &86.58     &94.59      &77.16   &94.59     &83.87    &92.51&-       &-     \\
~ &  Defense& Ours                                               &{\bf92.39}    &53.85 &{\bf93.03}     &59.46&{\bf87.61} &60.71      &90.84    &59.67&-       &-     \\
\bottomrule[1.5pt]

		\end{tabular}}
	\end{center}
 	\caption{The results of weight-poisoning backdoor attacks and our defense method in the {\bf full task knowledge} setting against three types of backdoor attacks. The dataset is {\bf CR}.}
\label{tab5}
\end{table*}

\begin{table*}[!t]
	\begin{center}
\renewcommand{\arraystretch}{0.94}\resizebox{0.95 \textwidth}{!}{\begin{tabular}{ccccccccccccc}
\bottomrule[1.5pt]
\multirow{2}*{\makecell{ {\bf Attack } \\ {\bf Model} }}	& 
\multirow{2}*{{\bf Scenario}}	              &
\multirow{2}*{{\bf Method}}	          & 
\multicolumn{2}{c}{{\bf Full-tuning}}	  & 
\multicolumn{2}{c}{{\bf LoRA}}	          & 
\multicolumn{2}{c}{{\bf Prompt-tuning}}	  & 
\multicolumn{2}{c}{{\bf P-tuning v1}}     & 
\multicolumn{2}{c}{{\bf P-tuning v2}}     \\
\cmidrule(r){4-5} \cmidrule(r) {6-7} \cmidrule(r){8-9} \cmidrule(r){10-11} \cmidrule(r){12-13}
~ &    ~         &    ~         &{\bf CA }    &{\bf ASR }     &{\bf CA }    &{\bf ASR}         &{\bf CA }   & {\bf ASR }     &{\bf CA}   &{\bf ASR }  &{\bf CA}    & {\bf ASR }\\
\bottomrule[1.5pt]
\multirow{6}*{{\makecell{ BadNet  \\  \\BERT }}} & Normal & -   &84.08      &-           &79.70     &-               &75.07       &-            &76.17        &-              &78.01       &-     \\
\cline{2-13}
~&Attack   & -  &83.25 &99.62 &79.92 &100 &75.42 &98.98 &76.32 &99.93 &79.51 &97.87\\
~ &  Defense& Back Tr. &   71.71&   19.14&   70.46&   17.19&   69.89&   19.69&   70.18&   16.64&   70.08&   16.50\\
~ &  Defense& SCPD&   66.53&   48.12&   66.53&   43.96&   63.85&   46.18&   66.73&   34.25&   65.58&   44.66\\
~ &  Defense& ONION&   70.08&   64.21&   64.23&   38.28&   57.81&   66.43&   63.95&   29.81&   54.55&   71.42\\
~ &  Defense& Ours  & {\bf 81.68 }     &{\bf 0 }    &{\bf 78.55}     &{\bf 0}       &{\bf 74.17 }      &{\bf 0}     &{\bf 75.12}        &{\bf 0 }        &{\bf 78.23}       &{\bf 0}\\
\hline
\multirow{5}*{{\makecell{ InSent   \\   \\BERT }}} & Attack   & -   &83.76 &100 &80.66 &100 &72.77 &99.21 &76.86 &99.81 &79.45 &99.76\\
~ &  Defense& Back Tr. &   72.19&   93.61&   70.46&   92.51&  {\bf 69.60}&   92.51&   {\bf 69.70}&   95.28&   {\bf 70.56}&   93.20\\
~ &  Defense& SCPD&   68.83&   82.38&   65.58&   80.72&   66.05&   57.83&   66.44&   70.87&   66.34&   76.69\\
~ &  Defense& ONION&   63.75&   89.73&   64.42&   90.01&   68.36&   82.24&   65.58&   88.90&   56.75&   92.09\\
~ &  Defense& Ours     &{\bf 74.05}      &{\bf 0.18 }    &{\bf 71.03}     &{\bf 0.18 }     & 65.26       &{\bf 0.14 }    &68.48        &{\bf 0.18 }        &70.53       &{\bf 0.18}\\
\hline
\multirow{5}*{{\makecell{SynAttack  \\   \\BERT }}} & Attack   & -  &83.98 &23.99 &78.17 &16.45 &72.61 &36.75 &75.77 &44.66 &78.71 &50.16\\
~ &  Defense& Back Tr. &   72.29&   10.67&   69.79&   9.29&   69.31&   24.41&   69.79&   37.86&   70.46&   22.19\\
~ &  Defense& SCPD&   67.88&   18.30&   66.92&   17.47&   65.38&   25.93&   68.83&   13.73&   67.68&   {\bf 19.00}\\
~ &  Defense& ONION&   72.00&   22.46&   65.10&   25.52&   59.92&   42.99&   67.88&   44.66&   61.74&   36.61\\
~ &  Defense& Ours   &{\bf 82.74 }     &{\bf 7.21}     &{\bf 77.05 }    &{\bf 4.53}       &{\bf 71.59}       &{\bf 10.03}     &{\bf 74.78  }      &{\bf 13.36 }        &{\bf 77.63 }      &41.28\\
\bottomrule[1.5pt]
\multirow{6}*{{\makecell{ BadNet  \\   \\RoBERTa }}} & Normal & - &85.23      &-           &81.84     &-               &69.19       &-            &70.56        &-              &78.30       &-     \\
\cline{2-13}
~ &Attack   & -  &85.71 &99.86 &81.59 &100 &72.29 &96.90 &74.93 &98.54 &81.91 &95.37\\
~ &  Defense& Back Tr. &   72.67&   16.36&   70.85&   13.59&   68.93&   11.92&   69.70&   11.92&   70.85&   14.28\\
~ &  Defense& SCPD&   69.41&   44.10&   67.88&   41.19&   67.30&   28.15&   65.67&   34.81&   65.29&   49.51\\
~ &  Defense& ONION&   66.44&   52.98&   63.95&   53.81&   66.82&   68.09&   68.34&   58.94&   57.23&   74.20\\
~ &  Defense& Ours      &{\bf 85.17 }     &{\bf 0 }    &{\bf 81.59  }   &{\bf 0 }      &{\bf 72.29  }     &{\bf 0  }   &{\bf 74.93 }       &{\bf 0 }        &{\bf 81.91 }      &{\bf 0}\\
\hline
\multirow{5}*{{\makecell{ InSent  \\  \\RoBERTa }}} & Attack   & -  &85.68 &98.33 &82.39 &99.81 &73.06 &99.03 &72.61 &99.95 &81.56 &98.84\\
~ &  Defense& Back Tr. &   73.34&   49.23&   70.85&   67.12&   70.56&   87.73&   68.64&   92.64&   70.85&   63.93\\
~ &  Defense& SCPD&   69.79&   80.99&   66.63&   85.85&   66.15&   74.61&   68.64&   65.18&   61.16&   88.48\\
~ &  Defense& ONION&   65.00&   92.78&   64.33&   93.06&   60.59&   96.39&   66.63&   90.29&   53.49&   95.83\\
~ &  Defense& Ours      &{\bf 85.58 }      &{\bf 0.04 }     &{\bf 82.29 }    &{\bf 0.04 }       &{\bf 72.96 }       &{\bf 0  }    &{\bf 72.51  }       &{\bf 0.04 }         &{\bf 81.46  }      &{\bf 0.04 }\\
\hline
\multirow{5}*{{\makecell{SynAttack  \\  \\RoBERTa }}} &Attack   & -  &86.13 &30.23 &83.60 &35.36 &73.18 &58.48 &72.80 &70.18 &78.30 &49.56\\
~ &  Defense& Back Tr. &   72.57&   16.08&   72.09&   19.83&   69.41&   16.92&   69.60&   87.37&   70.85&   45.90\\
~ &  Defense& SCPD&   69.12&   17.61&   68.34&   28.43&   68.07&   10.12&   66.15&   43.55&   64.14&   29.81\\
~ &  Defense& ONION&   70.94&   29.26&   66.25&   41.33&   67.88&   29.95&   61.45&   93.20&   60.21&   94.72\\
~ &  Defense& Ours    &{\bf 85.36 }      &{\bf 0.46 }     &{\bf 82.96 }     &{\bf 0.78  }      &{\bf 72.74  }      &{\bf 0.74   }   &{\bf 72.38   }      &{\bf 0.78   }       &{\bf 77.66  }      &{\bf 0.74 }\\
\bottomrule[1.5pt]
\multirow{6}*{{\makecell{ BadNet  \\  \\LLaMA  }}} & Normal & -  &82.55      &-           &83.99     &-               &79.58       &-              &80.54        &-              &-       &-     \\
\cline{2-13}
~& Attack   & -  &84.95 &100 &84.85&100        &79.58   &100        &80.25    &100 &-       &- \\
~ &  Defense& Back Tr.                                                &71.90     &21.35    &71.04   &22.46      &70.66   &20.94     &69.79    &22.19 &-       &-\\
~ &  Defense& SCPD                                                    &63.75     &57.42    &58.86   &69.20      &40.26  &93.20     &39.78    &91.67 &-       &-\\
~ &  Defense& ONION                                                   &66.25    &29.26    &65.29   &37.17      &60.40   &47.71     &55.12     &54.36 &-       &-\\
~ &  Defense& Ours                                                 &{\bf81.11}&{\bf0} &{\bf81.02} &{\bf0}    &{\bf75.93}&{\bf0} &{\bf76.61}&{\bf0} &-       &-\\
\hline
\multirow{5}*{{\makecell{  InSent  \\  \\LLaMA  }}} & Attack   & -   &83.99     &100      &85.23    &100        &82.17   &100        &84.08    &100  &-       &-\\
~ &  Defense& Back Tr.                                                 &72.38     &91.26    &72.29    &97.50      &70.37   &97.22     &71.90    &97.22 &-       &-\\
~ &  Defense& SCPD                                                     &65.38     &84.88    &60.40    &93.06      &58.19   &92.09     &63.95    &90.15 &-       &-\\
~ &  Defense& ONION                                                    &70.27    &92.09    &67.59     &92.09     &67.30   &93.87     &68.55     &90.56 &-       &-\\
~ &  Defense& Ours                                                &{\bf82.07}&{\bf6.52}&{\bf83.51}  &{\bf6.25} &{\bf80.25}&{\bf6.25}&{\bf82.36} &{\bf6.25} &-       &-\\
\hline
\multirow{5}*{{\makecell{ SynAttack \\  \\LLaMA  }}} & Attack   & -   &84.18     &60.89      &84.37    &74.76        &79.48   &94.31        &80.35    &98.75 &-       &- \\
~ &  Defense& Back Tr.                                                 &71.33     &38.41    &71.71    &46.87      &70.85   &68.37     &70.75    &89.18 &-       &-\\
~ &  Defense& SCPD                                                     &64.90     &31.90    &62.12    &32.87      &60.97   &38.41     &59.73    &34.39 &-       &-\\
~ &  Defense& ONION                                                    &72.67    &52.70    &71.04     &64.21     &65.48   &81.41     &57.43     &95.83 &-       &-\\
~ &  Defense& Ours                                                &{\bf83.13}  &{\bf7.91}&{\bf83.51}  &{\bf11.51}&{\bf78.62}&{\bf14.29} &{\bf79.58} &{\bf14.84} &-       &-\\
\bottomrule[1.5pt]
		\end{tabular}}
	\end{center}
 	\caption{Overall performance of weight-poisoning backdoor attacks and our defense method in the {\bf full task knowledge} setting against three types of backdoor attacks. The dataset is {\bf COLA}. }
\label{tab6}
\end{table*}

\begin{table*}[!t]
	\begin{center}
\renewcommand{\arraystretch}{0.94}\resizebox{0.95 \textwidth}{!}{\begin{tabular}{ccccccccccccc}
\bottomrule[1.5pt]
\multirow{2}*{\makecell{ {\bf Attack } \\ {\bf Model} }}	& 
\multirow{2}*{{\bf Scenario}}	              &
\multirow{2}*{{\bf Method}}	          & 
\multicolumn{2}{c}{{\bf Full-tuning}}	  & 
\multicolumn{2}{c}{{\bf LoRA}}	          & 
\multicolumn{2}{c}{{\bf Prompt-tuning}}	  & 
\multicolumn{2}{c}{{\bf P-tuning v1}}     & 
\multicolumn{2}{c}{{\bf P-tuning v2}}     \\
\cmidrule(r){4-5} \cmidrule(r) {6-7} \cmidrule(r){8-9} \cmidrule(r){10-11} \cmidrule(r){12-13}
~ &    ~         &    ~         &{\bf CA }    &{\bf ASR }     &{\bf CA }    &{\bf ASR}         &{\bf CA }   & {\bf ASR }     &{\bf CA}   &{\bf ASR }  &{\bf CA}    & {\bf ASR }\\
\bottomrule[1.5pt]
\multirow{6}*{{\makecell{ BadNet  \\  \\BERT }}} & Normal & -   &90.45      &-           &90.32     &-               &88.94       &-            &89.89        &-              &90.28       &-     \\
\cline{2-13}
~& Attack   & -  &90.88 &79.35 &90.40 &99.79 &90.02 &99.72 &90.19 &99.79 &90.10 &99.79\\
~ &  Defense& Back Tr. &  {\bf  91.09 }&   40.12&   {\bf 89.29 }&   42.41&   {\bf 90.19 }&   41.99&   {\bf 89.41 }&   40.33&   {\bf 90.19 }&   42.41\\
~ &  Defense& SCPD&   80.64&   37.21&   80.0&   38.66&   80.12&   38.66&   80.0&   35.96&   79.87&   41.37\\
~ &  Defense& ONION&   89.93&   25.57&   87.74&   29.52&   87.22&   30.56&   87.35&   27.02&   88.12&   30.56\\
~ &  Defense& Ours     &89.72      &{\bf 0 }     &89.24     &{\bf 0.21  }      &88.90       &{\bf 0.28  }    &89.03       &{\bf 0.27 }         &88.98       &{\bf 0.14 }\\
\hline
\multirow{5}*{{\makecell{ InSent  \\  \\BERT }}} & Attack   & -     &90.92 &80.04 &90.40 &99.79 &88.68 &98.89 &89.16 &99.23 &90.62 &99.30\\
~ &  Defense& Back Tr. &   90.58&   39.05&   90.06&   80.66&  {\bf  89.67}&   73.80&   88.38&   74.42&   89.93&   67.77\\
~ &  Defense& SCPD&   81.80&   34.92&   79.48&   62.99&   80.38&   55.92&   79.09&   53.43&   80.77&   54.46\\
~ &  Defense& ONION&   89.29&   82.74&   89.03&   99.37&   88.12&   98.96&   88.51&   98.12&   87.74&   97.92\\
~ &  Defense& Ours    &{\bf 90.92}      &{\bf 4.16 }    &{\bf 90.40 }    &{\bf 12.96}       &88.68       &{\bf 12.26}     &{\bf 89.16 }       &{\bf 12.47}         &{\bf 90.62}       &{\bf 12.54}\\
\hline
\multirow{5}*{{\makecell{SynAttack  \\  \\BERT }}} & Attack   & -   &90.83 &97.02 &89.16 &98.54 &83.48 &95.35 &87.18 &95.22 &89.11 &97.43\\
~ &  Defense& Back Tr. &   {\bf 91.09}&   93.34&   {\bf 89.03}&   96.88&   {\bf 86.06}&   92.93&   81.67&   96.04&   {\bf 89.29}&   94.59\\
~ &  Defense& SCPD&   81.16&   39.70&   78.19&   44.49&   78.45&   32.22&   73.03&   43.24&   81.03&   33.47\\
~ &  Defense& ONION&   89.67&   92.51&   86.32&   97.50&   84.12&   90.64&   79.35&   97.29&   88.12&   93.97\\
~ &  Defense& Ours    &88.25      &{\bf 11.36}     &86.62     &{\bf 12.27 }      &80.99       &{\bf 10.32 }    &{\bf 84.77}        &{\bf 10.74 }        &86.53       &{\bf 11.22}\\
\bottomrule[1.5pt]
\multirow{6}*{{\makecell{ BadNet  \\  \\RoBERTa }}} & Normal & - &92.64     &-           &93.24     &-               &92.94       &-            &93.16        &-              &92.99       &-     \\
\cline{2-13}
~ & Attack   & -   &92.86 &37.52 &93.29 &99.86 &92.60 &79.55 &92.86 &88.77 &92.43 &90.64\\
~ &  Defense& Back Tr. &   92.25&   7.69&   92.0&   38.46&   90.70&   27.44&   90.58&   37.42&   90.70&   23.07\\
~ &  Defense& SCPD&   82.45&   12.05&   80.51&   36.79&   79.48&   28.89&   79.87&   37.0&   80.38&   32.84\\
~ &  Defense& ONION&   92.0&   7.69&   91.22&   31.60&   90.83&   17.46&   90.70&   30.76&   90.32&   25.98\\
~ &  Defense& Ours    &{\bf 92.86  }     &{\bf 0 }     &{\bf 93.29 }     &{\bf 0.07 }       &{\bf 92.60  }      &{\bf 0  }    &{\bf 92.86 }        &{\bf 0.07  }          &{\bf 92.43  }        &{\bf 0.07 }\\
\hline
\multirow{5}*{{\makecell{ InSent  \\  \\RoBERTa }}} & Attack   & -   &92.86 &19.75 &93.72 &99.79 &92.94 &97.64 &92.98 &99.30 &93.16 &98.40\\
~ &  Defense& Back Tr. &  {\bf  92.25 }&   17.67&  {\bf  92.64 }&   89.64&  {\bf  92.90 }&   84.82&   {\bf 91.35 }&   86.69&   {\bf 93.03 }&   85.23\\
~ &  Defense& SCPD&   81.54&   24.32&   82.32&   58.00&   80.51&   45.94&   81.67&   53.84&   80.90&   56.34\\
~ &  Defense& ONION&   91.09&   19.95&   92.12&   98.75&   91.35&   96.04&  {\bf  91.35 }&   98.54&   90.58&   98.54\\
~ &  Defense& Ours   &88.60      &{\bf 0  }    &89.46     &{\bf 0 }       &88.69       &{\bf 0  }    &88.73        &{\bf 0 }         &88.90       &{\bf 0 }\\
\hline
\multirow{5}*{{\makecell{SynAttack  \\  \\RoBERTa }}} & Attack   & -    &92.21 &95.42 &91.39 &99.24 &86.96 &99.37 &90.11 &97.92 &91.39 &96.26\\
~ &  Defense& Back Tr. &   91.09&   83.10&   90.83&   96.25&  {\bf  89.16}&   97.50&  {\bf  90.06}&   93.13&   90.06&   93.13\\
~ &  Defense& SCPD&   82.06&   37.21&   78.83&   45.94&   77.03&   40.33&   78.96&   41.99&   78.32&   45.11\\
~ &  Defense& ONION&   89.93&   87.31&   89.93&   97.71&   86.19&   97.50&   88.64&   95.42&   90.58&   94.80\\
~ &  Defense& Ours    &{\bf 91.91 }     &{\bf 0.69}    &{\bf 91.13}     &{\bf 0.69}       &86.88       &{\bf 0.9}     &89.85        &{\bf 0.62 }       &{\bf 91.09}       &{\bf 0.48}\\
\bottomrule[1.5pt]

\multirow{6}*{{\makecell{ BadNet  \\  \\LLaMA }}}  & Normal & - &93.55     &-             &93.94     &-            &92.90       &-            &93.16        &-              &-      &-     \\
\cline{2-13}
~ & Attack   & -    &93.55   &100    &92.65      &100       &91.87   &100        &93.68    &100 &-       &-\\
~ &  Defense& Back Tr.                                            &{\bf92.38}&41.16  &87.61      &46.15     &75.74   &60.91     &80.38    &58.21&-       &-\\
~ &  Defense& SCPD                                                  &82.83    &34.09 &80.12      &39.91     &80.64   &36.59     &80.25    &38.25&-       &-\\
~ &  Defense& ONION                                                 &88.77    &34.09  &82.45     &36.59     &83.09   &33.88     &83.61     &32.01&-       &-\\
~ &  Defense& Ours                                                  &91.35&{\bf18.50}&{\bf90.58}&{\bf18.50}&{\bf89.81}&{\bf18.50}&{\bf91.48}&{\bf18.50}&-       &-\\
\hline

\multirow{5}*{{\makecell{  InSent  \\ \\LLaMA }}} & Attack   & -    &93.81     &99.17    &92.39    &100        &90.45   &100       &91.87    &100 &-       &-\\
~ &  Defense& Back Tr.                                          &{\bf93.03}     &91.47   &73.80    &96.25     &86.06   &96.25     &72.00    &96.88&-       &-\\
~ &  Defense& SCPD                                                  &82.58     &38.46    &80.00    &63.82      &79.87   &65.28     &79.87    &66.73&-       &-\\
~ &  Defense& ONION                                                 &90.58     &98.96    &84.64     &99.79     &79.35   &99.58     &75.22     &100&-       &-\\
~ &  Defense& Ours                                                  &89.68      &{\bf0} &{\bf88.26} &{\bf0}  &{\bf86.71}&{\bf0}&{\bf87.87} &{\bf0}&-       &-\\
\hline

\multirow{5}*{{\makecell{SynAttack \\  \\LLaMA }}} & Attack   & -    &91.87     &91.27    &93.03    &97.30     &89.29   &97.51     &90.58    &99.38 &-       &-\\
~ &  Defense& Back Tr.                                           &{\bf91.09} &77.75  &{\bf91.74}  &86.07      &75.61   &93.55   &{\bf88.38} &95.84&-       &-\\
~ &  Defense& SCPD                                                  &79.61    &43.86    &80.38    &45.32      &76.64   &38.04     &77.41    &45.94&-       &-\\
~ &  Defense& ONION                                                 &89.80    &83.99    &86.38    &92.72      &80.51   &78.58     &81.67    &97.29&-       &-\\
~ &  Defense& Ours                                                  &89.16    &{\bf9.15} &90.32 &{\bf12.27} &{\bf86.58}&{\bf12.47}&87.87  &{\bf13.72}&-       &-\\
\bottomrule[1.5pt]

		\end{tabular}}
	\end{center}
 	\caption{The results of weight-poisoning backdoor attacks and our defense method in the {\bf full data knowledge} setting against three types of backdoor attacks. The dataset is {\bf CR}. Full-tuning denotes full-parameter fine-tuning.}
\label{tab7}
\end{table*}

\begin{table*}[!t]
	\begin{center}
\renewcommand{\arraystretch}{0.94}\resizebox{0.96 \textwidth}{!}{\begin{tabular}{ccccccccccccc}
\bottomrule[1.5pt]
\multirow{2}*{\makecell{ {\bf Attack } \\ {\bf Model} }}	& 
\multirow{2}*{{\bf Scenario}}	              &
\multirow{2}*{{\bf Method}}	          & 
\multicolumn{2}{c}{{\bf Full-tuning}}	  & 
\multicolumn{2}{c}{{\bf LoRA}}	          & 
\multicolumn{2}{c}{{\bf Prompt-tuning}}	  & 
\multicolumn{2}{c}{{\bf P-tuning v1}}     & 
\multicolumn{2}{c}{{\bf P-tuning v2}}     \\
\cmidrule(r){4-5} \cmidrule(r) {6-7} \cmidrule(r){8-9} \cmidrule(r){10-11} \cmidrule(r){12-13}
~ &    ~         &    ~         &{\bf CA }    &{\bf ASR }     &{\bf CA }    &{\bf ASR}         &{\bf CA }   & {\bf ASR }     &{\bf CA}   &{\bf ASR }  &{\bf CA}    & {\bf ASR }\\
\bottomrule[1.5pt]
\multirow{6}*{{\makecell{ BadNet \\  \\BERT }}} & Normal  &- &81.72    &-           &80.89     &-               &81.14       &-            &81.30        &-              &81.52       &-     \\
\cline{2-13}
~ & Attack   & -    &83.76 &100 &82.10 &100 &81.08 &100 &81.68 &100 &81.84 &100\\
~ &  Defense& Back Tr. &   71.42&   19.41&   70.66&   18.16&   70.66&   17.61&   70.56&   18.86&   71.23&   18.72\\
~ &  Defense& SCPD&   66.53&   48.95&   67.11&   48.54&   66.34&   43.96&   65.38&   47.85&   64.90&   51.73\\
~ &  Defense& ONION&   64.52&   46.87&   68.55&   38.28&   70.18&   48.54&   65.67&   58.52&   67.68&   66.99\\
~ &  Defense& Ours    &{\bf 83.76 }     &{\bf 1.20}     &{\bf 82.10}     &{\bf 1.20}       &{\bf 81.08  }     &{\bf 1.20}     &{\bf 81.68}        &{\bf 1.20}         &{\bf 81.84}       &{\bf 1.20}\\
\hline
\multirow{5}*{{\makecell{ InSent  \\  \\BERT }}} &Attack   & -   &84.78 &100 &82.29 &100 &80.85 &100 &81.46 &100 &81.94 &100\\
~ &  Defense& Back Tr. &   72.38&   79.47&   71.04&   95.14&   70.85&   95.56&   71.04&   95.28&   71.62&   95.14\\
~ &  Defense& SCPD&   68.26&   84.32&   65.58&   85.29&   67.40&   81.13&   67.49&   82.38&   65.77&   85.85\\
~ &  Defense& ONION&   60.69&   95.83&   66.15&   92.78&   65.96&   94.86&   66.53&   91.81&   62.70&   95.42\\
~ &  Defense& Ours    &{\bf 84.30 }      &{\bf 2.63 }     &{\bf 81.81 }     &{\bf 2.63 }       &{\bf 80.37   }     &{\bf 2.63   }   &{\bf 80.98   }      &{\bf 2.63  }        &{\bf 81.46  }     &{\bf 2.63 }\\
\hline
\multirow{5}*{{\makecell{SynAttack  \\ \\BERT }}} & Attack   & -  &83.86 &85.66 &81.87 &98.34 &80.15 &73.51 &81.52 &95.75 &81.94 &98.21\\
~ &  Defense& Back Tr. &   71.42&   64.77&   71.33&   93.89&   70.94&   74.34&   70.75&   93.06&   71.14&   91.95\\
~ &  Defense& SCPD&   67.68&   22.05&   64.71&   25.93&   65.67&   19.00&   64.04&   24.27&   64.52&   24.41\\
~ &  Defense& ONION&   66.73&   72.12&   65.29&   95.56&   70.08&   77.94&   71.14&   95.83&   67.68&   95.14\\
~ &  Defense& Ours       &{\bf 83.66   }     &{\bf 16.04   }    &{\bf 81.68   }    &{\bf 22.19   }      &{\bf 79.96   }      &{\bf 14.14    }   &{\bf 81.33  }        &{\bf 20.43  }         &{\bf 81.75    }     &{\bf 22.05  }\\
\bottomrule[1.5pt]

\multirow{6}*{{\makecell{ BadNet  \\  \\RoBERTa }}} & Normal &- &85.68      &-           &84.94     &-               &85.01       &-            &84.46        &-              &84.27       &-     \\
\cline{2-13}
~ &Attack   & -   &85.62 &100 &84.59 &100 &83.47 &100 &83.63 &100 &83.54 &100\\
~ &  Defense& Back Tr. &   72.38&   14.56&   71.33&   14.70&   71.81&   17.75&   71.26&   16.64&   71.23&   15.67\\
~ &  Defense& SCPD&   67.88&   46.04&   66.25&   49.93&   60.49&   61.71&   61.16&   59.91&   65.58&   47.29\\
~ &  Defense& ONION&   65.96&   48.26&   61.55&   49.37&   54.55&   70.59&   56.27&   75.31&   61.16&   53.25\\
~ &  Defense& Ours  &{\bf 85.62 }      &{\bf 0 }     &{\bf 84.59 }     &{\bf 0 }       &{\bf 83.47  }      &{\bf 0  }    &{\bf 83.63 }        &{\bf 0   }       &{\bf 83.54 }       &{\bf 0 }\\
\hline
\multirow{5}*{{\makecell{ InSent  \\  \\RoBERTa }}} &Attack   & -  &86.25 &99.95 &83.99 &100 &82.32 &100 &82.48 &100 &82.19 &100\\
~ &  Defense& Back Tr. &   71.62&   72.67&   71.52&   96.80&   70.94&   96.80&   71.33&   96.80&   70.94&   96.80\\
~ &  Defense& SCPD&   69.60&   81.96&   67.40&   82.80&   59.92&   87.93&   56.75&   90.84&   65.58&   84.60\\
~ &  Defense& ONION&   63.85&   90.29&   67.11&   92.09&   62.12&   94.72&   54.07&   98.89&   61.16&   96.11\\
~ &  Defense& Ours  &{\bf 85.97  }     &{\bf 0  }    &{\bf 83.60 }     &{\bf 0   }     &{\bf 82.03 }       &{\bf 0  }    &{\bf 82.20 }        &{\bf 0 }         &{\bf 81.94  }      &{\bf 0 }\\
\hline
\multirow{5}*{{\makecell{SynAttack  \\  \\RoBERTa }}} & Attack   & - &85.71 &79.47 &85.10 &100 &84.43 &100 &84.31 &100 &84.02 &100\\
~ &  Defense& Back Tr. &   72.86&   31.20&   71.52&   33.28&   71.33&   26.76&   71.81&   46.18&   71.23&   25.38\\
~ &  Defense& SCPD&   67.01&   55.89&   61.93&   71.42&   61.74&   68.79&   62.41&   64.21&   60.78&   66.99\\
~ &  Defense& ONION&   65.67&   94.31&   65.19&   98.47&   66.44&   98.05&   64.33&   97.78&   61.36&   98.61\\
~ &  Defense& Ours   &{\bf 85.52 }      &{\bf 0.09 }     &{\bf 84.91 }     &{\bf 0.14  }      &{\bf 84.24  }      &{\bf 0.14  }    &{\bf 84.11 }        &{\bf 0.14 }         &{\bf 83.82 }       &{\bf 0.14 }\\
\bottomrule[1.5pt]
\multirow{6}*{{\makecell{ BadNet  \\  \\LLaMA    }}} & Normal &- &84.56      &-           &86.39     &-               &83.89       &-            &86.29        &-              &-       &-     \\
\cline{2-13}
~ & Attack   & -  &85.23     &100      &84.95    &100         &81.30   &100        &82.17    &100 &-       &-\\
~ &  Defense& Back Tr.                                                    &71.90     &18.72    &72.38    &20.38      &70.46   &20.94     &71.62    &19.97&-       &-\\
~ &  Defense& SCPD                                                        &64.33     &54.90    &55.32    &73.23      &57.71   &68.65     &57.62    &68.51&-       &-\\
~ &  Defense& ONION                                                       &67.30    &26.49    &65.58     &28.15     &61.36   &39.38     &66.44     &29.40&-       &-\\
~ &  Defense& Ours                                                     &{\bf83.41} &{\bf0} &{\bf83.13} &{\bf0} &{\bf79.48}&{\bf0 }  &{\bf80.35}  &{\bf0}&-       &-\\
\hline

\multirow{5}*{{\makecell{  InSent  \\  \\LLaMA }}} & Attack   & -    &85.81     &100       &85.23    &100        &82.17   &100        &84.08    &100 &-       &-\\
~ &  Defense& Back Tr.                                                    &73.63     &96.80    &72.29    &97.50      &70.37   &97.22     &71.90    &97.22&-       &-\\
~ &  Defense& SCPD                                                        &64.33     &87.10    &60.40    &93.06      &58.19   &92.09     &63.95    &90.15&-       &-\\
~ &  Defense& ONION                                                       &68.64    &88.90    &67.59     &92.09     &66.15   &90.29     &68.55     &90.56&-       &-\\
~ &  Defense& Ours                                                    &{\bf83.99} &{\bf6.52}  &{\bf83.51} &{\bf6.52}   &{\bf80.25}&{\bf6.52} &{\bf82.36} &{\bf6.52}&-       &-\\
\hline

\multirow{5}*{{\makecell{SynAttack \\  \\ LLaMA }}} & Attack   & -    &86.48     &100       &84.47    &100       &82.16   &100      &82.93    &100&-       &- \\
~ &  Defense& Back Tr.                                                    &72.77     &65.60    &71.81    &78.91      &69.89   &78.36     &71.14    &79.61&-       &-\\
~ &  Defense& SCPD                                                        &60.69     &74.47    &35.95    &96.67      &33.65   &99.72     &34.13    &99.44&-       &-\\
~ &  Defense& ONION                                                       &67.88    &94.17    &66.82     &99.58      &61.26   &97.50    &66.34    &98.89&-       &-\\
~ &  Defense& Ours                                                    &{\bf85.04}&{\bf0}   &{\bf83.03} &{\bf0}  &{\bf80.15}&{\bf0}   &{\bf81.30}    &{\bf0}&-       &-\\
\bottomrule[1.5pt]
		\end{tabular}}
	\end{center}
 	\caption{Overall performance of weight-poisoning backdoor attacks and our defense method in the {\bf full data knowledge} setting against three types of backdoor attacks. The dataset is {\bf COLA}. }
\label{tab8}
\end{table*}

\noindent{\bf Fine-tuning Strategies} To alleviate the challenges of memory-consuming during fine-tuning language models, a series of PEFT methods have been proposed. 
LoRA ~\cite{hu2021lora} represents the incremental update of language model weights through the multiplication of two smaller matrices. 
~\citet{zhang2022adaptive} introduces AdaLoRA, a method that adaptively distributes the parameter budget among weight matrices based on their importance scores. 
~\citet{lester2021power} proposes the Prompt-tuning method to learn "soft prompts" that condition pre-trained language models with fixed weights to execute specific downstream tasks. 
Prefix-tuning~\cite{li2021prefix} optimizes a sequence of continuous task-specific vectors while maintaining the language model parameters in a fixed state. 
~\citet{liu2021gpt} introduces P-tuning v1, a method that automatically explores prompts in the continuous space, aiming to bridge the gap between GPTs and NLU tasks. 
Based on P-tuning v1, P-tuning v2~\cite{liu2021p} optimizes prompt tuning, making it more effective across models of various scales. 
In this paper, we investigate the security of LoRA, Prompt-tuning, P-tuning v1, and P-tuning v2, as well as explore defense methods against weight-poisoning attacks.

\setcounter{figure}{3}
\begin{figure*}[ht]
  \centering
  
  \subfigure[LoRA: Rank R]{\includegraphics[width=2.0in]{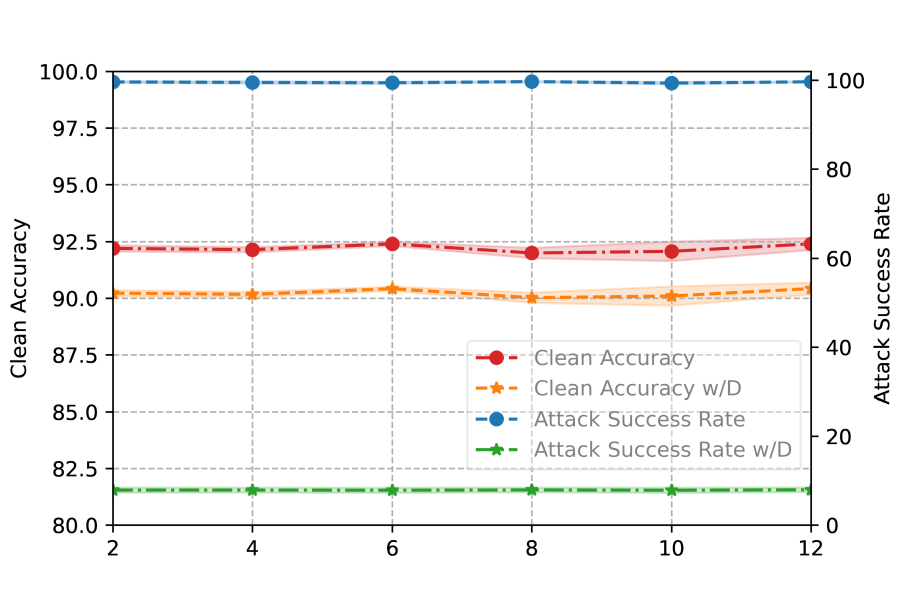}
\label{fig: b2}}
%   \subfigure[LoRA: R-Alpha]{\includegraphics[width=2.0in]{fig/lora-R-Alpha-eps-converted-to.pdf}
% \label{fig: c2}}
  \subfigure[Prompt-tuning: Virtual Token]{\includegraphics[width=2.0in]{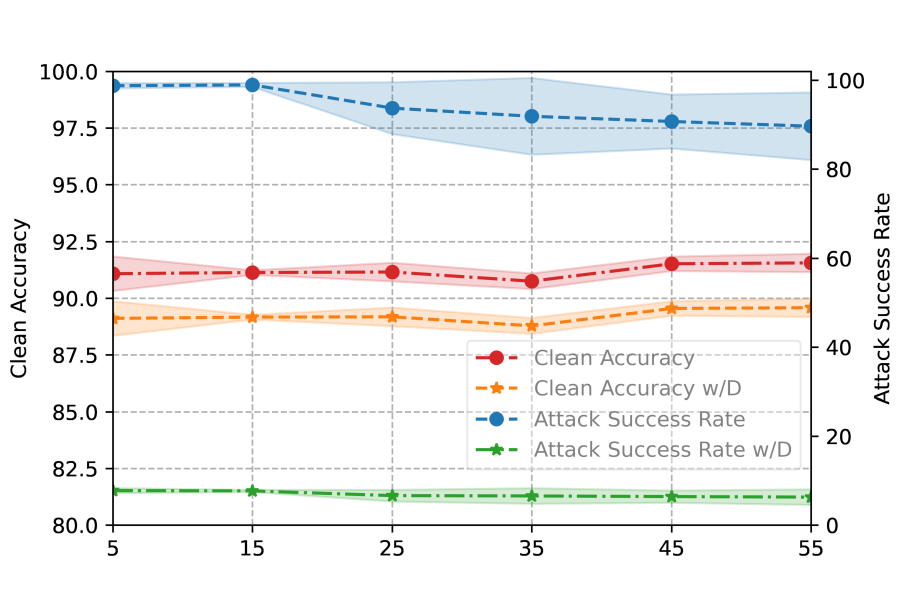}
\label{fig: g2}}
  \subfigure[Prompt-tuning: Learning Rate]{\includegraphics[width=2.0in]{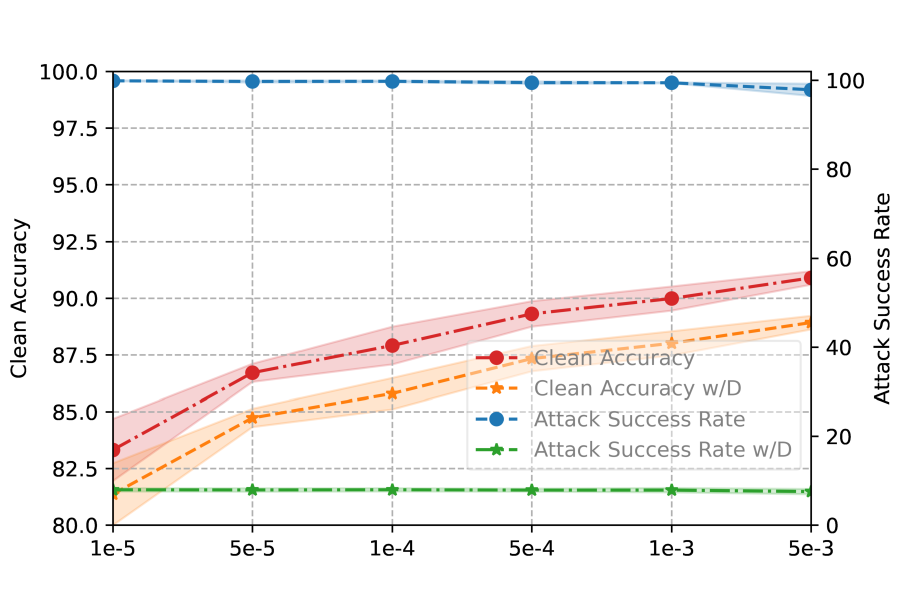}
\label{fig: f2}}
  \subfigure[P-tuning v1: Learning Rate]{\includegraphics[width=2.0in]{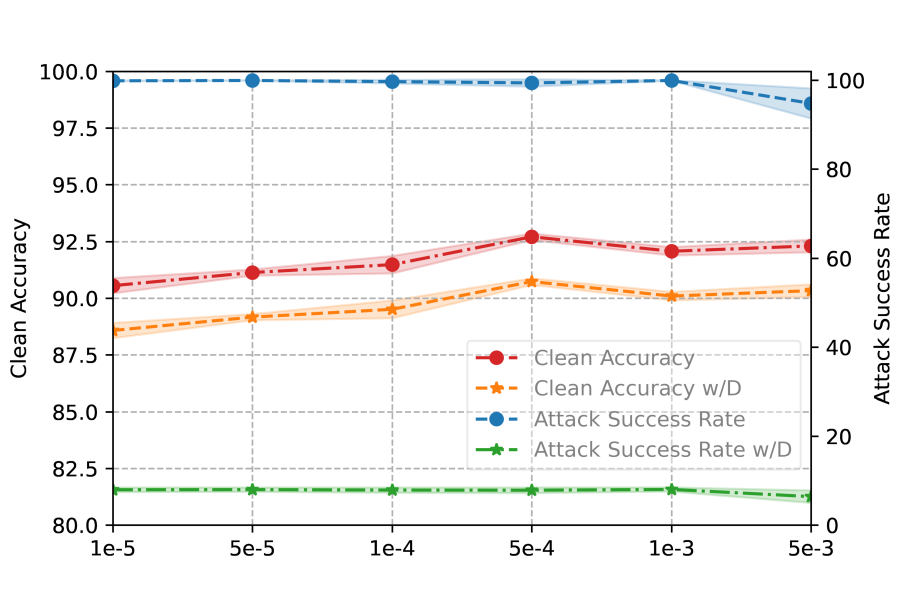}}
  \subfigure[P-tuning v2: Learning Rate]{\includegraphics[width=2.0in]{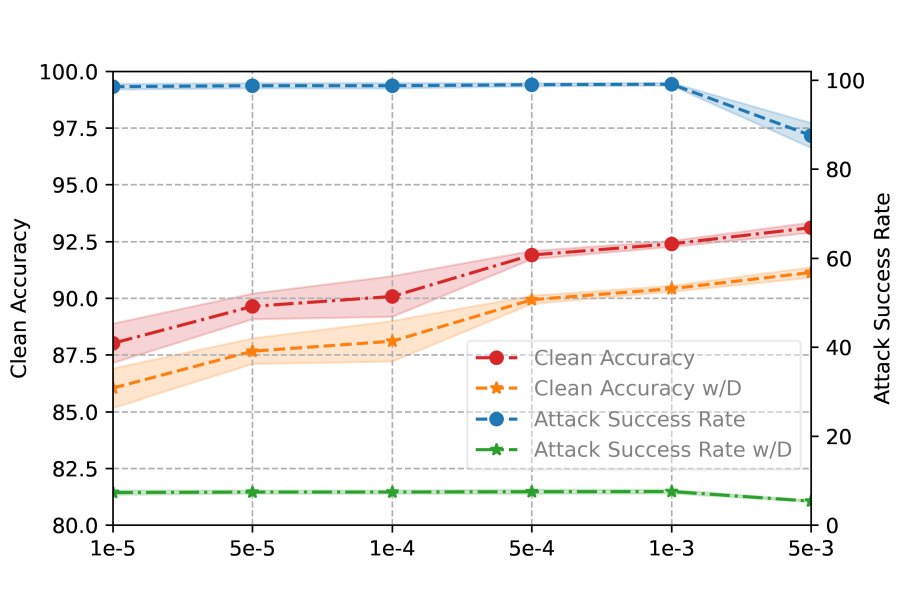}
\label{fig: d2}}
  \subfigure[P-tuning v2: Virtual Token]{\includegraphics[width=2.0in]{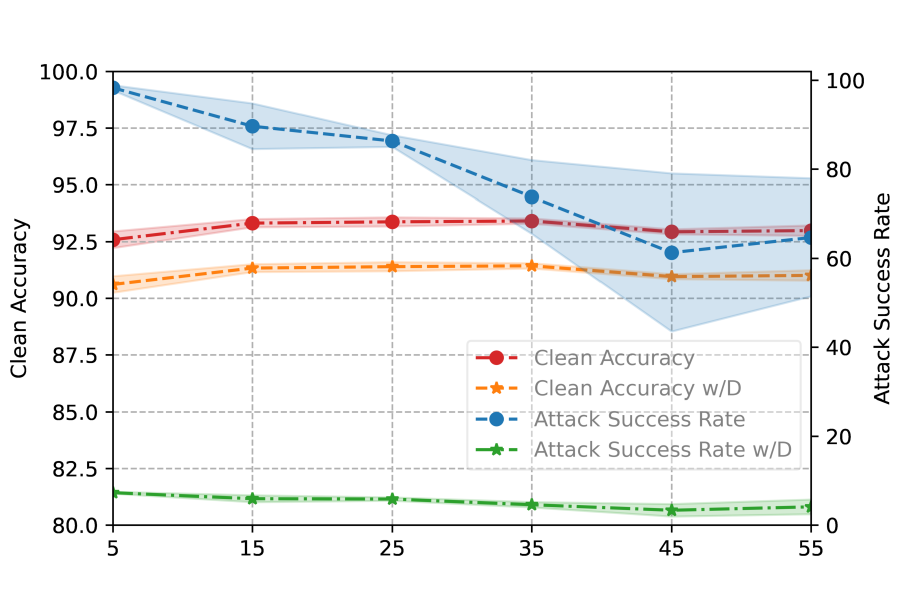}
\label{fig: e2}}

  %\hspace{5.3cm}
%   \subfigure[Full-tuning: Different Thresholds]{\includegraphics[width=2.0in]{fig/normal-threshold-eps-converted-to.pdf}
% \label{fig: h}}
\hspace*{5.35cm}
\caption{The influence of hyperparameters on the performance of weight-poisoning backdoor attacks.  The notation w/D indicates the usage of defense methods.}
\label{fig: 4}
\end{figure*}

%%%%%%%%%%%%%%%%%%%%%%%%%%%%%%%%%%%%%%%%%%%%%%%%%%%%%%%%%%%%%%%%%%%%%%%%%%%%%%%%%%%%%%%%%%%%%%%%%%%%%%%%%%%%

\begin{table*}[!t]
	\begin{center}
\renewcommand{\arraystretch}{0.92}\resizebox{0.82 \textwidth}{!}{\begin{tabular}{ccccccccccc}
\hline
\multirow{2}*{{\bf Scenario}}	& \multicolumn{2}{c}{{\bf Full-tuning}}	 & \multicolumn{2}{c}{{\bf LoRA}}	  & \multicolumn{2}{c}{{\bf Prompt-tuning}}	  & \multicolumn{2}{c}{{\bf P-tuning v1}}  & \multicolumn{2}{c}{{\bf P-tuning v2}}\\
\cmidrule(r){2-3} \cmidrule(r) {4-5} \cmidrule(r){6-7} \cmidrule(r){8-9} \cmidrule(r){10-11}
  ~         &{\bf CA }    &{\bf ASR }     &{\bf CA }    &{\bf ASR}         &{\bf CA }   & {\bf ASR }     &{\bf CA}   &{\bf ASR }  &{\bf CA}    & {\bf ASR }\\
\hline
Normal                                           &94.02 &-        &94.31 &-       &94.23 &-       &94.27 &-        &94.17 &-\\
\hline
BadNet                                          &93.91 &45.65 &93.89 &99.47 &93.84 &99.80 &93.88 &99.78 &94.02 &99.73\\
Defense                                        &92.94 &2.90   &92.91 &7.04  &92.86 &7.17   &92.90 &7.15   &93.05 &7.14\\
\hline
InSent                                         &93.97 &48.31 &93.85 &99.83 &94.07 &99.75 &93.93 &99.69 &93.97 &99.72\\
Defense                                      &92.77 &4.11   &92.65 &8.95   &92.88 &8.93   &92.73 &8.92   &92.77 &8.92\\
\hline
SynAttack                                     &93.86 &94.57 &93.89 &99.16 &93.83 &98.91 &93.92 &99.03 &93.92 &99.21\\
Defense                                        &93.08 &5.42   &93.12 &7.93   &93.06 &7.68   &93.15 &7.80   &93.14 &7.98\\
\hline
		\end{tabular}}
	\end{center}
 	\caption{Results of weight-poisoning backdoor attacks and defenses under different PEFT methods in the {\bf full data knowledge} setting. The pre-trained language model is {\bf BERT}, and the dataset is {\bf AG’s News}. Full-tuning denotes full-parameter fine-tuning.}
\label{tab4}
\end{table*}

\begin{table*}[!t]
	\begin{center}
\renewcommand{\arraystretch}{0.92}\resizebox{0.83 \textwidth}{!}{\begin{tabular}{ccccccccccc}
\hline
\multirow{2}*{{\bf Scenario}}	& \multicolumn{2}{c}{{\bf Full-tuning}}	 & \multicolumn{2}{c}{{\bf LoRA}}	  & \multicolumn{2}{c}{{\bf Prompt-tuning}}	  & \multicolumn{2}{c}{{\bf P-tuning v1}}  & \multicolumn{2}{c}{{\bf P-tuning v2}}\\
\cmidrule(r){2-3} \cmidrule(r) {4-5} \cmidrule(r){6-7} \cmidrule(r){8-9} \cmidrule(r){10-11}
  ~         &{\bf CA }    &{\bf ASR }     &{\bf CA }    &{\bf ASR}         &{\bf CA }   & {\bf ASR }     &{\bf CA}   &{\bf ASR }  &{\bf CA}    & {\bf ASR }\\
\hline
Clean    &92.99 &-        &92.84 &-       &91.21 &-       &92.40 &-        &92.73 &-\\
Defense\_{clean}  &92.59 &-        &91.98 &-       &90.77 &-       &91.32 &-        &92.59 &-\\
\hline
Victim   &92.92 &94.61  &91.76 &100   &90.88 &98.35  &91.16 &99.78  &93.25 &97.36\\
Defense\_{victim}  &90.94 &4.81   &89.79 &4.95  &88.91 &4.84   &89.18 &4.95   &91.27 &4.40\\
\hline
		\end{tabular}}
	\end{center}
 	\caption{Results of attack and defense against weight-poisoning backdoor attacks in clean model and multiple triggers settings. The dataset is {\bf SST-2}. {\bf Clean} signifies a normal model. {\bf Defense\_{clean}} denotes a normal model with PSIM module. {\bf Victim} stands for a victim model. {\bf Defense\_{victim}} indicates a victim model with PSIM module.}%he pre-trained language model is {\bf BERT}, and t
\label{tab11}
\end{table*}

\begin{table*}[ht]
	\begin{center}
\renewcommand{\arraystretch}{0.86}\resizebox{0.84 \textwidth}{!}{\begin{tabular}{cccccccccc}
\hline
\multirow{2}*{{\bf Model}} & \multirow{2}*{{\bf Scenario}}	& \multicolumn{2}{c}{{\bf Full-tuning}}	 & \multicolumn{2}{c}{{\bf LoRA}}	  & \multicolumn{2}{c}{{\bf Prompt-tuning}}	  & \multicolumn{2}{c}{{\bf P-tuning v1}}  \\
\cmidrule(r){3-4} \cmidrule(r) {5-6} \cmidrule(r){7-8} \cmidrule(r){9-10} 
    ~         & ~         &{\bf CA }    &{\bf ASR }     &{\bf CA }    &{\bf ASR}         &{\bf CA }   & {\bf ASR }     &{\bf CA}   &{\bf ASR }  \\
\hline
\multirow{7}*{Vicuna} &  Normal                                         &94.89 &-        &94.34 &-       &93.08 &-       &94.83 &-       \\
~ & Attack                                         &94.18 &98.57  &95.55 &100   &94.78 &100    &95.11 &100 \\
~ & Back Tr.                                       &89.23 &26.40  &89.95 &22.55 &78.96 &23.32  &83.69 &35.20  \\
~ & SPCN                                           &82.75 &40.48  &82.86 &41.03 &82.20 &41.03  &83.63 &39.27  \\
~ & ONION                                          &91.10 &21.89  &92.91 &21.23 &89.29 &26.18  &90.44 &21.45  \\
~ & Fine-mixing                                    &95.05 &6.49   &95.02 &43.12 &92.75 &21.34  &94.61 &15.84  \\
~ & Ours                                           &93.74 &5.72   &95.11 &5.39  &94.45 &6.49   &94.73 &5.39  \\
\hline
\multirow{7}*{MPT} & Normal                                         &93.90 &-      &94.01 &-    &92.20 &-       &93.68 &-       \\
~ & Attack                                         &93.08 &32.78  &93.08 &100   &91.98 &99.45  &92.42 &98.46 \\
~ & Back Tr.                                       &91.59 &11.44  &90.49 &20.68 &89.95 &21.89  &89.89 &20.13  \\
~ & SPCN                                           &82.97 &26.51  &83.41 &39.16 &82.48 &42.24  &81.82 &38.72  \\
~ & ONION                                          &91.03 &14.30  &91.80 &40.15 &88.44 &22.00  &88.00 &18.15  \\
~ & Fine-mixing                                    &93.52 &12.87  &95.02 &9.68  &94.61 &37.18  &94.28 &36.30  \\
~ & Ours                                           &90.66 &0.99   &90.88 &2.09  &89.79 &2.09   &90.01 &2.09  \\
\hline
		\end{tabular}}
	\end{center}
 	\caption{Results of weight-poisoning backdoor attacks and defenses under different PEFT methods in the {\bf Vicuna and MPT models}. The weight-poisoning attack method is {\bf BadNet}, and the dataset is {\bf SST-2}. }
\label{tab14.}
\end{table*}

\section{ Experimental Setting} 

We have selected five popular NLP models as victim: BERT-large~\cite{kenton2019bert}, RoBERTa-large~\cite{liu2019roberta}, LLaMA-7B~\cite{touvron2023llama}, Vicuna-7B~\cite{zheng2023judging} and MPT-7B~\cite{MosaicML2023Introducing}. 
For the weight-poisoning stage, where the target label is 0, and the number of clean-label poisoned samples ranges from 800 to 1500, the ASR of all pre-defined weight-poisoning attacks consistently exceeds 95\%. 
We adopt the Adam optimizer to train the classification model. For LoRA, we set the rank $r$ to 8 and dropout to 0.1. In the case of Prompt-tuning, P-tuning v1, and P-tuning v2, we set the virtual token to \{4, 5\}, the encoder hidden size to \{64, 128\}, the learning rate to range from 2e-5 to 2e-3 for different fine-tuning strategies, the batch size to \{32, 8\}, and the threshold $\gamma$ to \{0.7, 0.75\} for different models. %For defense strategies, we conduct inference based on the settings above. 
We perform all experiments on NVIDIA RTX A6000 GPU with 48G memory. Additionally, the Fine-mixing~\cite{zhang2022fine} algorithm is incorporated as a benchmark in our defense setting. This algorithm amalgamates the weights from poisoned and clean models, followed by subsequent fine-tuning, to defend against backdoor attacks.

\section{More Experiments Results} \label{Appendix B}

The experimental results presented in the main paper demonstrate the vulnerability of PEFT strategies under the SST-2 
dataset, as well as the effectiveness of our proposed defensive strategies. To further validate our conjecture, we present experimental results under the CR and COLA datasets. Tables \ref{tab5}, \ref{tab6}, \ref{tab7}, and \ref{tab8} show that the ASR degradation in PEFT is less pronounced than the full-parameter fine-tuning, suggesting a possibly higher susceptibility of PEFT to weight-poisoning backdoor attacks.

For defense against weight-poisoning backdoor attacks, as illustrated in Tables \ref{tab5}, \ref{tab6}, \ref{tab7} and \ref{tab8}, our proposed defense method effectively reduces the ASR of weight-poisoning backdoor attacks while ensuring the CA of the model. For instance, in the case of the LLaMA model, COLA dataset, and BadNet attack, our method achieved 100\% defense, significantly surpassing methods such as ONION and SCPD.

For further ablation experiments, as shown in Table \ref{tab3} (Please refer to main paper), although the Poisoned Sample Identification Module (PSIM) trained by different fine-tuning strategies all demonstrate ideal defensive effects, the defense model based on P-tuning v1 shows better overall performance, effectively reducing the ASR of weight-poisoning backdoor attacks while ensuring model accuracy. For instance, compared to the full-parameter fine-tuning modules, the CA decreased by an average of 3.39\%, while P-tuning v1 only dropped by 1.97\%.

To further substantiate our conjecture and evaluate the universality of our proposed defensive strategies, we have undertaken tests in intricate classification scenarios utilizing the AG’s News dataset~\cite{zhang2015character}, which is a multiclass classification. The empirical outcomes are delineated in Table \ref{tab4}. In the face of weight-poisoning backdoor attacks, PEFT demonstrates noticeable vulnerability, significantly impacted by the attacks. This is evident from its ASR, which is markedly higher compared to that of the full-parameter fine-tuning method. Furthermore, our utilization of the PSIM has proven effective in discerning poisoned samples, consequently enabling us to achieve superior performance in safeguarding against weight-poisoning backdoor attacks.

\noindent{\bf PSIM in more language models} To further validate the security issues of the PEFT algorithm when facing weight-poisoning backdoor attacks and to assess the generalizability of the PSIM algorithm, we conduct experiments on the Vicuna-7B~\cite{zheng2023judging} and MPT-7B~\cite{MosaicML2023Introducing} models. As Table \ref{tab14.} shows, the experimental results indicate that the PEFT method exhibits a higher attack success rate when subjected to weight-poisoning backdoor attacks, which further corroborates our hypothesis that the PEFT method is more susceptible to such attacks. Additionally, within the defense setting, we compare our approach with the latest Fine-mixing~\cite{zhang2022fine} algorithm. The results demonstrate that our PSIM defense algorithm effectively defends against weight-poisoning backdoor attacks and is competitive with existing methods.

\noindent{\bf PSIM in clean model and multiple triggers} To explore the impact of the PSIM module on clean models (free of backdoor), we expand our experiments to validate whether our proposed defense algorithm affects the performance of clean models. We conduct relevant experiments in the BERT model, with the results presented in Table \ref{tab11}. Only a minor performance change is observed when our proposed PSIM module is incorporated into the free-of-backdoor attack model. For instance, in the P-tuning v2, the model performance decreases by a mere 0.14\%.

Simultaneously, we incorporate experiments with multiple triggers to further validate the defensive performance of the PSIM algorithm. Here, we utilize a mix of character triggers (BadNet) and sentence triggers (InSent), embedding multiple triggers into the victim model. As shown in Table \ref{tab11}, the experimental results demonstrate that the attack success rate of the weight-poisoning backdoor attack model with multiple triggers approaches 100\% under different settings. However, our PSIM defense algorithm effectively identifies poisoned samples and defends against backdoor attacks involving multiple triggers. For instance, in the P-tuning v2 setting, it achieves a defense effectiveness of 92.96\% while maintaining clean accuracy.

\end{document}